\documentclass{aa}  

\usepackage{graphicx}
\usepackage{txfonts}
\usepackage{hyperref}
\usepackage{adjustbox}
\begin{document} 

%def's
\def\teff{${T}_{\rm eff}$}
\def\kms{{km\,s}$^{-1}$}
\def\logg{$\log g$}
\def\micro{$\xi_{\rm t}$}
\def\macro{$\zeta_{\rm RT}$}
\def\rad{$v_{\rm r}$}
\def\vsini{$v\sin i$}
\def\ebv{$E(B-V)$}
\def\kepler{\textit{Kepler}}

   \title{The $Kepler$ view of magnetic chemically peculiar stars}

\author{S.~H{\"u}mmerich\inst{1,2}
        \and Z.~Mikul\'{a}\v{s}ek\inst{3}
        \and E.~Paunzen\inst{3}
        \and K.~Bernhard\inst{1,2}
        \and J.~Jan{\'i}k\inst{3}
        \and I.A.~Yakunin\inst{4}
        \and T.~Pribulla\inst{5}
        \and M.~Va\v{n}ko\inst{5}
        \and L.~Mat\v{e}chov\'a\inst{3}
        }
\institute{Bundesdeutsche Arbeitsgemeinschaft f{\"u}r Ver{\"a}nderliche Sterne e.V. (BAV), D-12169 Berlin, Germany, \email{ernham@rz-online.de}
\and American Association of Variable Star Observers (AAVSO), 49 Bay State Rd, Cambridge, MA 02138, USA
\and Department of Theoretical Physics and Astrophysics, Masaryk University, Kotl\'a\v{r}sk\'a 2, 611\,37 Brno, Czech Republic
\and Special astrophysical observatory of the RAS, Nizhnii Arkhyz, Karachai-Cherkessian Republic, 369167, Russia 
\and Astronomical Institute, Slovak Academy of Sciences, 059 60 Tatransk{\'a} Lomnica, Slovakia}

\date{}

% \abstract{}{}{}{}{} 
% 5 {} token are mandatory
 
  \abstract
   {Magnetic chemically peculiar (mCP) stars exhibit complex atmospheres that allow the investigation of such diverse phenomena as atomic diffusion, magnetic fields, and stellar rotation. The advent of space-based photometry provides the opportunity for the first precise characterizations of the photometric variability properties of these stars, which might advance our understanding of the processes involved and the atmospheric structures of mCP stars.}
   {We carried out a search for new mCP stars in the \kepler\ field with the ultimate aim of investigating their photometric variability properties using \kepler\ data. As an aside, we describe criteria for selecting mCP star candidates based on light curve properties, and assess the accuracy of the spectral classifications provided by the MKCLASS code.}
   {As only very few known mCP stars are situated in the \kepler\ field, we had to depend largely on alternative (non-spectroscopic) means of identifying suitable candidates that rely mostly on light curve properties; in particular we relied on monoperiodic variability and light curve stability. Newly acquired and archival spectra were used to confirm most of our mCP star candidates. Linear ephemeris parameters and effective amplitudes were derived from detrended \kepler\ data.}
   {Our final sample consists of 41 spectroscopically confirmed mCP stars of which 39 are new discoveries, 5 candidate mCP stars, and 7 stars in which no chemical peculiarities could be established. Our targets populate the whole age range from zero-age main sequence to terminal-age main sequence and are distributed in the mass interval from 1.5\,$M_{\odot}$ to 4\,$M_{\odot}$. About 25\,\% of the mCP stars show a hitherto unobserved wealth of detail in their light curves indicative of complex surface structures. We identified light curve stability as a primary criterion for identifying mCP star candidates among early-type stars in large photometric surveys, and prove the reliability of the spectral classifications provided by the MKCLASS code.}
        {}


   \maketitle
%
%-------------------------------------------------------------------

\section{Introduction} \label{introduction}

Chemically peculiar (CP) stars make up about 10\,\% of early-B to early-F main-sequence stars. They are characterized by enhanced (or weakened) absorption lines that indicate peculiar surface abundances \citep{preston74}. The origin of the chemical peculiarities is generally explained by atomic diffusion, i.e. the competition between radiative levitation and gravitational settling \citep{michaud70,richer00}. While most elements tend to sink under the force of gravity, those with numerous absorption lines near the local flux maximum are driven outwards by radiative pressure. This mechanism can efficiently work in CP stars because of their slow rotation rates (absence of meridional flows) and calm radiative outer layers. Table \ref{table_CPs} presents an overview of the main subgroups of CP stars and some of their characteristics. Further subgroups have been subsequently defined, such as $\lambda$ Bootis stars \citep{parenago58,paunzen04}, but these are not relevant to the present investigation.

\begin{table*}
\caption{Main subgroups of CP stars according to \citet{preston74}.}
\label{table_CPs}
\begin{center}
\begin{tabular}{lll}
\hline
\hline 
ID & Common name & Main characteristics \\
\hline
CP1 & metallic-line (Am) stars & underabundances of Ca and Sc, overabundances of iron-peak and heavier elements \\
CP2 & magnetic Ap stars & enhanced Si, Cr, Sr, Eu, or rare earth elements \\
CP3 & HgMn stars & enhanced Hg, Mn, and other heavy elements\\
CP4 & He-weak stars & weak He lines \\
\hline
\end{tabular}
\end{center}
\end{table*}

The CP2/4 stars are in the focus of the present investigation. They possess globally organized magnetic fields with strengths of up to several tens of kG \citep{auriere07}, which are likely of fossil origin \citep[e.g.][]{braithwaite04}.\footnote{For convenience, these objects are referred to hereafter as magnetic chemically peculiar (mCP) stars.} 
The mCP stars exhibit a non-uniform distribution of chemical elements and have spots of enhanced or depleted element abundance \citep{michaud81}, which can be studied in detail using Doppler imaging (DI; \citealt{kochukhov17}). Flux is redistributed in these spots through bound-free \citep{peterson70,lanz96} and bound-bound transitions (\citealt{wolff71,molnar73}; cf. also \citealt{krticka13}). 

Magnetic and rotational axes are commonly unaligned in mCP stars (oblique rotator model; \citealt{stibbs50}), and -- as the star rotates -- the changing viewing angle results in strictly periodic spectroscopic and photometric variations with the rotation period. Stars exhibiting this kind of variability are traditionally referred to as $\alpha^2$ Canum Venaticorum (ACV) variables \citep{GCVS}. A subset of CP2 stars exhibit pulsational variability with very small amplitudes in the period range from 5 m to 20 m (high-overtone, low-degree, and non-radial pulsation modes). These are known as rapidly oscillating Ap stars \citep{kurtz82}, also referred to as roAp stars. In contrast, non-pulsating CP2 stars are sometimes referred to as noAp stars.

Most mCP stars exhibit rotational periods of the order of several days, but there exists a population of very slow rotators with periods of months, years \citep[e.g.][]{wolff75}, or even centuries \citep{mathys17}; this represents a significant percentage of mCP stars \citep{mathys16}.

Several studies have explored the photometric variability of mCP stars and it has been shown that rotational light curves can be adequately described by a sine wave and its first harmonic \citep[e.g.][]{north84,mathys85,mikulasek07b,bernhard15a,huemmerich16}. This finding has generally been interpreted in terms of a rather simple spot geometry, in which the distribution of chemical under- and overabundances reflects the symmetries of the magnetic structure.

The long, ultra-precise and well-sampled time series data provided by space missions, such as the \textit{CoRoT} \citep{Corot} or \kepler\ \citep{Kepler} satellites, have revolutionized the field of variable star research and in particular asteroseismology. The detection limit for variability signals has decreased dramatically, which has enabled the first detailed investigations and classifications of variable stars with very low amplitudes and has revealed unprecedented detail and complexity in their light curves. 

Unfortunately, as yet, only a few (noAp) mCP stars have been investigated in detail using space photometry \citep{lueftinger10,paunzen15,escorza16,drury17}, and only a handful of known mCP stars are situated in the \kepler\ field. Nevertheless, some of these studies already hinted at the occurrence of more complex rotational light changes in mCP stars \citep{paunzen15}. Accurate light curves containing maximum information content are a necessary basis for any modelling attempts \citep{prvak15}, thus a larger sample size of highly accurate mCP star light curves is desirable.

To improve this situation, we carried out a search for new mCP stars in the \kepler\ field with the ultimate aim of studying their rotational light variability properties using \kepler\ data. In addition, we describe criteria for selecting mCP star candidates based on light curve properties and provide an independent assessment of the accuracy of spectral classifications provided by the MKCLASS code \citep{gray16}. Data sources, target selection criteria, and further aspects of methodology are described in Sect. \ref{dataso}. Results are presented in Sect. \ref{result} and discussed in Sect. \ref{discussion}. We conclude in Sect. \ref{conclusion}.

%-----------------------------------------------------------------------

\section{Data sources and methodology} \label{dataso}

%-----------------------------------------------------------------------

\subsection{\kepler\ observations} \label{kepler}

The prime \kepler\ mission had been operating from 2009 May 2 until 2013 May 8, when the loss of a second reaction wheel on the spacecraft necessitated its abortion. Its main goal was the detection of planets around other stars via the transit method to determine the frequency of Earth-like planets in or near the habitable zone of Sun-like stars \citep{Kepler}. The long, quasi-uninterrupted and high-quality time series data have also been used for stellar variability analysis and in particular asteroseismology. This allows us to probe the interior of stars and the determination of properties such as radius and age through observations of modes of oscillation.

The \kepler\ spacecraft is equipped with a differential photometer with a 115 square-degree field of view in the constellations Cygnus and Lyra,  known as the Kepler field, which allows the continuous monitoring of $\sim$150\,000 stars. It boasts an aperture of 0.95\,m and its detectors consist of 21 modules each equipped with two 2200x1024 pixel CCDs. To optimize solar panel efficiency, the spacecraft completed a 90 degree roll every three months. Therefore, \kepler\ observations are divided into four quarters each year. \kepler\ provides single-passband light curves of micro-magnitude precision. The mission has been very successful in its main goal and has discovered thousands of exoplanet candidates. For more information on the \kepler\ spacecraft, we refer to \citet{Kepler} and \citet{koch10}.

\kepler\ data were acquired at two different time resolutions: the long-cadence (LC), 29.4 m sampling mode \citep{LCdata} and the short-cadence (SC), $\sim$1 m sampling mode \citep{SCdata}. We only employed  LC data in the investigation of our final sample stars.

\kepler\ Simple Aperture Photometry (SAP) data, which undergo only basic calibration, show strong instrumental effects that are caused by issues such as pointing drift or the return to operating focus and temperature after a quarterly roll. These trends are not easily distinguished from genuine stellar signals. The Presearch Data Conditioning (PDC) module of the data processing pipeline aims at removing signatures related to systematic error sources from the light curves \citep{LCdata,murphy14,stumpe14}. Several algorithms were employed, of which the most recent  (applying to LC PDC light curves formatted as v5.0 and greater) is the  multi-scape MAP (msMAP) approach \citep{stumpe14}, which is a wavelet-based band-splitting framework for removing systematics from the light curves. For more information on instrumental effects in \kepler\ data, see the Kepler Data Characteristics Handbook\footnote{https://archive.stsci.edu/kepler/manuals/Data\_Characteristics.pdf}.

The quality of the PDC-processed data constitute a major improvement from SAP data. However, instrumental trends of varying amplitude and origin still remain in the data, such as the small exponential thermal recoveries seen at the downlink events that affect the zero points of the data at the beginning of a quarter. Further detrending is necessary to eliminate, or at least diminish in strength, these trends (cf. Sect. \ref{dataan}).

%-----------------------------------------------------------------------

\subsection{Target selection criteria} \label{target}

In the following, the methods and selection criteria that have been employed in the construction of the present sample of stars are described.

\subsubsection{Known mCP stars in the \kepler\ field of view} \label{known_mCPs}

The most comprehensive catalogue of CP stars is the collection of \citet{RM09}, which contains more than 3\,500 mCP stars or mCP star candidates. Objects in this catalogue are not explicitly subdivided according to the classification established by \citet{preston74}. Therefore, we resorted to the listed spectral types to distinguish between the different groups of CP stars (mainly denoted as "Si", "Sr", "SrEuSi", "He weak", "HgMn", and so on). As has been pointed out (Sect. \ref{introduction}), only very few confirmed or candidate mCP stars are situated in the \kepler\ field, most of which have already been the subject of an investigation. Table \ref{table1} gives basic information on these objects.

KIC\,2569073 is not included in the \citet{RM09} catalogue but has been subsequently identified as an mCP star by \citet{drury17}. The star KIC\,8351193, which is missing from Table \ref{table1}, was found to show strong enhancement in Si by \citet{tkachenko13} and has consequently been listed as a chemically peculiar rotating variable with a spectral type of B9 V Si by \citet{balona15}. However, subsequent studies have not confirmed the Si peculiarity in this star \citep{niemczura15}, which is a low-metallicity object showing an almost featureless spectrum and rotates at 162\,\kms. It is certainly not a classical mCP star and has not been considered for the present investigation.\footnote{Please note that while \citet{tkachenko13} comment on the strong Si abundance of KIC\,8351193 in the text of their paper, the corresponding table (Table 3) in the same work lists a Si abundance of -1.20 dex, as compared to the solar abundance. Perhaps, then, an error in sign is at the root of this discrepancy.}

\begin{table}
\caption{Basic information on the known confirmed or candidate mCP stars in the \kepler\ field of view. If not indicated otherwise, spectral types were extracted from \citet{RM09}.}
\label{table1}
\begin{center}
\begin{adjustbox}{max width=0.5\textwidth}
\begin{tabular}{llll}
\hline
\hline 
KIC & AltID & Spec.type & Remark \\
\hline
6020867 & HD 175132                      & B9 Si      &  suspected variable{$^{1}$}\\
4136285 & HD 176582        & B5 He wk.  &  V545\,Lyr; spect.,\\
                    &                                                                    &                                               &  mag. and phot. var.{$^{2}$} \\
8677585 & TYC 3132-1679-1  & A5 EuCr   &  roAp star{$^{3}$} \\ 
9020199 & HD 182895        & F2 CrEu?   &  $\delta$\,Scuti star{$^{4}$}\\
3246460 & HD 185330        & B5 He wk.  &  HR 7467 \\                    
9851142 & HD 188854        & A5-F0 Sr   &  V2094\,Cyg; ecl.bin.\\
                                &                                                                        &                                               &  and $\gamma$\,Dor star{$^{5}$}\\
8324268 & HD 189160A       & B9 Si      &  V2095\,Cyg; ACV{$^{6}$}\\         
2569073 & UCAC4 640-066878 & A5 SrCrEu{$^{7}$} &  V694\,Lyr; ACV{$^{7}$}\\
\hline
\multicolumn{4}{l}{$^{1}$ \citet{koen02} $^{2}$ \citet{bohlender11}} \\
\multicolumn{4}{l}{$^{3}$ \citet{balona13} $^{4}$ \citet{uytterhoeven11}} \\
\multicolumn{4}{l}{$^{5}$ \citet{cakirli15} $^{6}$ \citet{balona11}} \\
\multicolumn{4}{l}{$^{7}$ \citet{drury17}} \\
\end{tabular}
\end{adjustbox}  
\end{center}
\end{table}

Of the stars listed in Table \ref{table1}, only KIC\,4136285 and KIC\,8324268 were included in our sample. The other objects have been rejected for several reasons.

KIC\,9851142 (V2094\,Cyg) has been classified as A5p Sr \citep{cannon23} and A7p CrEu \citep{zirin51} in the older literature, which suggests an mCP star. It has consequently been listed as an ACV variable in the General Catalogue of Variable Stars (GCVS; \citealt{GCVS}). However, more recent studies have shown that this is in fact a CP1 star \citep[e.g.][]{grenier99,cakirli15} and a $\gamma$ Doradus pulsator \citep{cakirli15}. In agreement with this classification, no rotationally modulated light variations are present in the light curve of this star, which has been excluded from further investigation.

KIC\,9020199 has been classified as spectral type F0p by \citet{macrae52} and F2p CrEu? by \citet{kharchenko01}. According to the \citet{RM09} catalogue, however, the peculiarity type classification is doubtful. Using the spectrum from the LAMOST-\kepler\ project (cf. Section \ref{LAMOST_MKCLASS}), we find that the star exhibits the signature of a CP1 star. Furthermore, \kepler\ SC data indicate the presence of multiple frequencies in the range between 6 d$^{-1}$ and 20\,d$^{-1}$; the most prominent peaks are at 7.078\,d$^{-1}$, 7.482\,d$^{-1}$, and 8.381\,d$^{-1}$. The pulsational properties and light curve of the star are in agreement with a $\delta$ Scuti classification, and it has been classified as such by \citet{uytterhoeven11}. We find no evidence for rotationally modulated variations in the \kepler\ light curve.

The CP4 star KIC\,3246460 shows complex and multiperiodic light variations that cannot be explained by rotation (alone). While this object is of considerable interest, its variability pattern is not representative of the class of mCP stars and has therefore been excluded from the present investigation. KIC\,6020867 was only observed during quarter Q12; the resulting time baseline of $\sim$83 days was deemed insufficient for an in-depth analysis of its light variations. Data for KIC\,2569073 are not publically available but have been extracted from the superstamp images around NGC 6791 by \citet{drury17}, who investigated the light variability of this star in detail. Finally, the roAp star KIC\,8677585 has been thoroughly investigated elsewhere \citep[e.g.][]{balona13}.

\subsubsection{Sample selection criteria} \label{sample_selection}

In view of the dearth of known mCP stars in the \kepler\ field, we had to resort to other means of identification. During our former searches for new ACV variables in other sky survey data, we systematically investigated unclassified early-type (late-B and early-A) variables to search for the presence of ACV variables among these objects \citep[e.g.][]{bernhard15a}. Promising candidates were selected by their photometric variability pattern, i.e. period, amplitude, light curve shape/stability, and monoperiodic variability; an investigation of their Fourier amplitude spectra; and available catalogue data, such as spectral type and colour indices. This approach, which has proved useful, is discussed in more detail below and has been adopted in a slightly modified form for the creation of an initial candidate sample for the present study.

As it is not straightforward to distinguish between variability induced by rotation and other sources such as pulsation or orbital motion, several aspects need to be taken into account. It has been shown that, generally, the rotational light curves of ACV variables can be adequately described by a sine wave and its first harmonic (cf. Sect. \ref{introduction}). In orientations in which two spots of optically active elements come into view during a single rotation cycle, the resulting light curve is a double wave \citep{maitzen80}. This often results in characteristic light curves that are tell-tale signs of ACV variables. However, in case of two bright spots of similar extent and photometric properties, the resulting "maxima" would be approximately the same height and the light curve would resemble, and could be easily confused with, that of a non-eclipsing close binary star;  these stars are also termed, rather confusingly, ellipsoidal variables. From standard, ground-based single-passband photometry alone, it is difficult to distinguish these light curves from double-humped ACV variables.

The situation is different when dealing with sufficiently precise photometric data, such as provided by space-bound missions. We were successful in computing phenomenological models of binary light curves outside eclipses that fit the observed light curves with an accuracy of 0.0003\,mag \citep{mikulasek17}. The obtained models of unspotted ellipsoidal variables provide a useful instrument for distinguishing ellipsoidal variables from double-waved ACV variables.\footnote{The full description of this procedure and range of our results, together with the simulation of light curves of non-eclipsing close binaries containing spotted components, is in preparation and will be published elsewhere. The assumption of an ellipsoidal variable with a spotted component might explain the variability of the non-mCP stars from our list that display a significant long-term development of their light curves (cf. Sects. \ref{periodogram} and \ref{nonCPs}).} We  applied this procedure to all candidate mCP stars with stable light curves and concluded that their light curves cannot be interpreted as the light curves of ellipsoidal variables.

In the spectral range of our sample stars, several kinds of pulsating variables are found \citep{GCVS,catelan15}.  $\beta$ Cephei (BCEP) variables occupy the range from $\sim$O8--B6 and exhibit periods between about 0.1--0.6\,d. Slowly pulsating B (SPB) stars are encountered roughly between spectral types B2--B9 and vary on timescales of about 0.4 to 5\,d. Finally,  $\gamma$ Doradus (GDOR) stars are $\sim$A7--F7 objects with periods of the order of 0.3 to 3\,d. In period and amplitude space, these stars (in particular SPB and GDOR variables) may occupy similar positions as ACV variables that usually show periods in excess of 0.5\,d with a peak distribution at around 2\,d \citep{RM09,bernhard15a}.

\begin{figure}
        \includegraphics[width=\columnwidth]{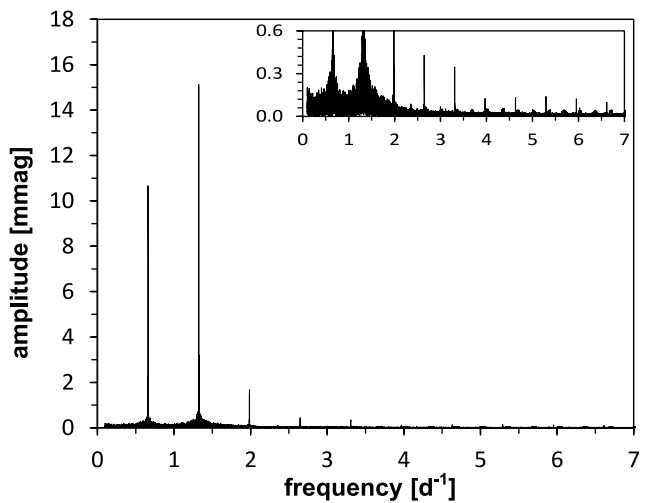}
    \caption{Fourier amplitude spectrum of the ACV variable KIC\,6950556 (spectral type A0 V Si) based on non-detrended Kepler PDC flux. The spectrum is indicative of a single independent frequency and its corresponding harmonics. The inset shows a zoomed-in version with the main peaks off scale, highlighting the corresponding harmonics; axes are the same as in the main plot.}
    \label{fig_ampf}
\end{figure}

An important means of distinguishing between types of variable stars is an investigation of their Fourier amplitude spectra. Except for roAp stars, which are rare and special objects exhibiting oscillations in the period range from 5 m to 20 m, ACV variables are characterized by a strictly monoperiodic variability that remains stable over long periods of time. Therefore ACV frequency spectra are typically characterized by a single independent frequency and its corresponding harmonics, which are a consequence of localized spots and characteristic of rotating variables \citep{balona15}. As an example, the frequency spectrum of the ACV variable KIC\,6950556 is provided in Fig. \ref{fig_ampf}.

On the other hand, early-type pulsators, such as the SPB or GDOR stars, are notoriously multiperiodic. With the advent of ultra-precise space photometry, a wealth of information on the pulsational properties of these objects has been accumulated, and the emerging complex picture has challenged the "coarse" variability groups based on ground-based observations. Multiple frequencies are routinely found, for  example for SPB variables \citep{balona11,mcnamara12}, which often also exhibit periods in the $\beta$ Cephei star realm in which case they are referred to as SPB/$\beta$ Cep hybrid pulsators. In fact, the presence of multiple frequencies is commonly used as a criterion to distinguish pulsating stars such as SPB stars from rotating variables \citep[e.g.][]{mcnamara12}. In summary, the presence of multiple independent modes has been established to be an important criterion for distinguishing pulsators from rotating variables.

The above-listed considerations have led to the development of the following selection criteria:
\begin{itemize}
        \item[a)] spectral type between late-B and early-F, corresponding colour index or effective temperature
        \item[b)] period longer than 0.5\,d        \item[c)] the presence of a single independent variability frequency and corresponding harmonics        \item[d)] stable or marginally changing light curve throughout the covered time span        \item[e)] variability amplitude of several hundredths of a magnitude (cf. \citealt{bernhard15a}) or less\end{itemize}

To create a sample of candidate mCP stars, we applied these criteria to the  \kepler\ light curves of early-type stars chosen via their colour indices or effective temperatures as listed in \citet{mathur17}.\footnote{Some of the stars investigated in this way were taken from a candidate list kindly provided by Prof. Don Kurtz (personal communication).} Stars selected in this way are referred to hereafter as photometric candidates. In regard to item c), we allowed for the presence of slow variations of the base period that might be indicative of a binary companion. More information concerning the application of this item is provided in Sect. \ref{periodogram}. Furthermore, bearing in mind the high precision of the \kepler\ photometry, item d) has been used as a rough guideline only, and stars exhibiting much smaller variability amplitudes were allowed.

These criteria were not applied strictly but used as a guideline in the identification of suitable candidates. Generally, in order to maintain a time-efficient approach, we only checked data from selected quarters at this stage for a given object.

It has to be noted that, obviously, we did not carry out an exhaustive search among all early-type targets in the \kepler\ field. Rather, for the present study, a sample of about 50 stars was collected, which was deemed a good compromise between an economic approach and a sufficiently large sample size to estimate the suitability of the chosen selection criteria and pursue the goals of the present study.

We also note that the employed method is bound to miss very long period ACV variables. In regard to item d), it has been deemed necessary that at least four variability cycles have been covered to estimate the stability of the light curve. Given that \kepler\ data cover a time span of about four years, this places an upper limit of about one year on the variability period. Furthermore, Kepler data are known to suffer from significant low-frequency noise (cf. Sections \ref{kepler} and \ref{ephemerides}), which renders the identification of low-frequency signals doubtful \citep[e.g.][]{balona11}. Therefore, periods longer than $\sim$10 days need to be regarded with some caution.

All in all, 53 stars (51 photometric candidates and the well-established mCP stars KIC\,4136285 and\,KIC 8324268) entered into the final sample. Literature information on our target stars was collected from the SIMBAD \citep{SIMBAD} and VizieR \citep{VIZIER} databases. Newly-acquired and archival spectroscopic observations were also included into the analysis and are described in the following sections.

%-----------------------------------------------------------------------

\subsection{Spectroscopic observations} \label{spectr}

This section provides information on the newly acquired spectra, employed instrumentation, and archival spectroscopic observations and classifications that were employed to investigate our sample stars in more detail.

\subsubsection{Spectroscopic observations at SAO RAS}

Spectroscopic observations were obtained using the UAGS spectrograph\footnote{http://www.sao.ru/Doc-en/Telescopes/small/instrum.html} mounted at the Cassegrain focus of the 1 m Zeiss-1000 telescope of the Special Astrophysical Observatory (SAO) of the Russian Academy of Science (RAS) during two observational runs in May and December 2016. The log of observations is provided in Table \ref{table_spectra_RAS}. The spectra were obtained using the diffraction grating R1302 with 1302 lines/mm and an EEV-42-40 CCD with 2048x2028 pixels. A He+Ne+Ar source of line spectrum was used for wavelength calibration. The data were obtained with spectral resolution R $\approx$ 2500 in the wavelength range 4192–5590 \AA. The typical signal-to-noise (S/N) ratio of the spectra is S/N$\approx$100. Standard primary reduction of the 2D spectra was performed using the Image Reduction and Analysis Facility (IRAF) software package (noao.twodspec package). A sample spectrum is shown in Fig. \ref{fig_spectra}.

\begin{table}
\caption{Log of spectroscopic observations obtained with the Zeiss-1000 telescope of the SAO RAS.}
\label{table_spectra_RAS}
\begin{center}
\begin{adjustbox}{max width=0.5\textwidth}
\begin{tabular}{cccc}
\hline
\hline 
KIC & mag(V) & T\_exp(s) & S/N \\
\hline
\multicolumn{4}{l}{May 25/26, 2016} \\
\hline
KIC\,10324412 & 8.2 & 600 & 130 \\
KIC\,6864569 & 10.0 & 1800 & 115 \\
KIC\,6065699 & 7.7 & 600 & 190 \\
KIC\,5473826 & 10.9 & 2400 & 100 \\
\hline
\multicolumn{4}{l}{May 27/28, 2016} \\
\hline
KIC\,6278403 & 8.7 & 1200 & 115 \\
KIC\,6206125 & 10.6 & 2400 & 53 \\
\hline
\multicolumn{4}{l}{Dec 08/09, 2016} \\
\hline
KIC\,10281890 & 10.4 & 3000 & 250 \\
KIC\,11671226 & 11 & 3600 & 200 \\
KIC\,8324268 & 7.9 & 600 & 250 \\
KIC\,7778838 & 11.6 & 3600 & 60 \\
\hline
\end{tabular}
\end{adjustbox}  
\end{center}
\end{table}

\subsubsection{Spectroscopic observations at Star{\'a} Lesn{\'a} Observatory}

Additional spectra were obtained with the {\'e}chelle eShel spectrograph of the Star{\'a} Lesn{\'a} Observatory (High Tatras, Slovakia), which is part of the Astronomical Institute of the Slovak Academy of Sciences. The spectrograph is attached to a 60 cm (f/12.5) Zeiss reflecting telescope (G1 pavilion). The employed CCD camera (ATIK 460EX) uses a 2750x2200 chip resulting in a resolution between 11\,000 and 12\,000 within a spectral range 4150 to 7600 \AA. The reduction of the raw frames and extraction of the 1D spectra using the IRAF package tasks, Linux shell scripts, and FORTRAN programs, have been described by \citet{pribulla15}. The wavelength reference system, as defined by the preceding and following Th-Ar exposures, was stable to within 0.1\,$\mathrm{km\,s}^{-1}$.

\subsubsection{LAMOST-\kepler\ project and the MKCLASS code} \label{LAMOST_MKCLASS}

The aim of the LAMOST-\kepler\ project is to collect high-quality low-resolution spectra of stars in the \kepler\ field with the LAMOST telescope (\citealt{lamost1,lamost2}). In this framework, \citet{gray16} have presented MK spectral classifications of more than 100,000 LAMOST spectra of 80,447 objects situated in the \kepler\ field. To this end, the automatic classification code MKCLASS \citep{MKCLASS} was employed, which is a program designed to emulate the steps used in the traditional classification process by an expert human classifier. First results have shown that, given input spectra of sufficient S/N, the results of MKCLASS compare well with human classifiers (precision of 0.6 spectral subclass and half a luminosity class; \citealt{MKCLASS}). An important feature of MKCLASS is its ability to identify traditional chemical peculiarities, such as are commonly found in the CP1 and CP2 stars. For more information on the MKCLASS code, we refer to \citet{MKCLASS}.

\begin{figure}
        \includegraphics[width=\columnwidth]{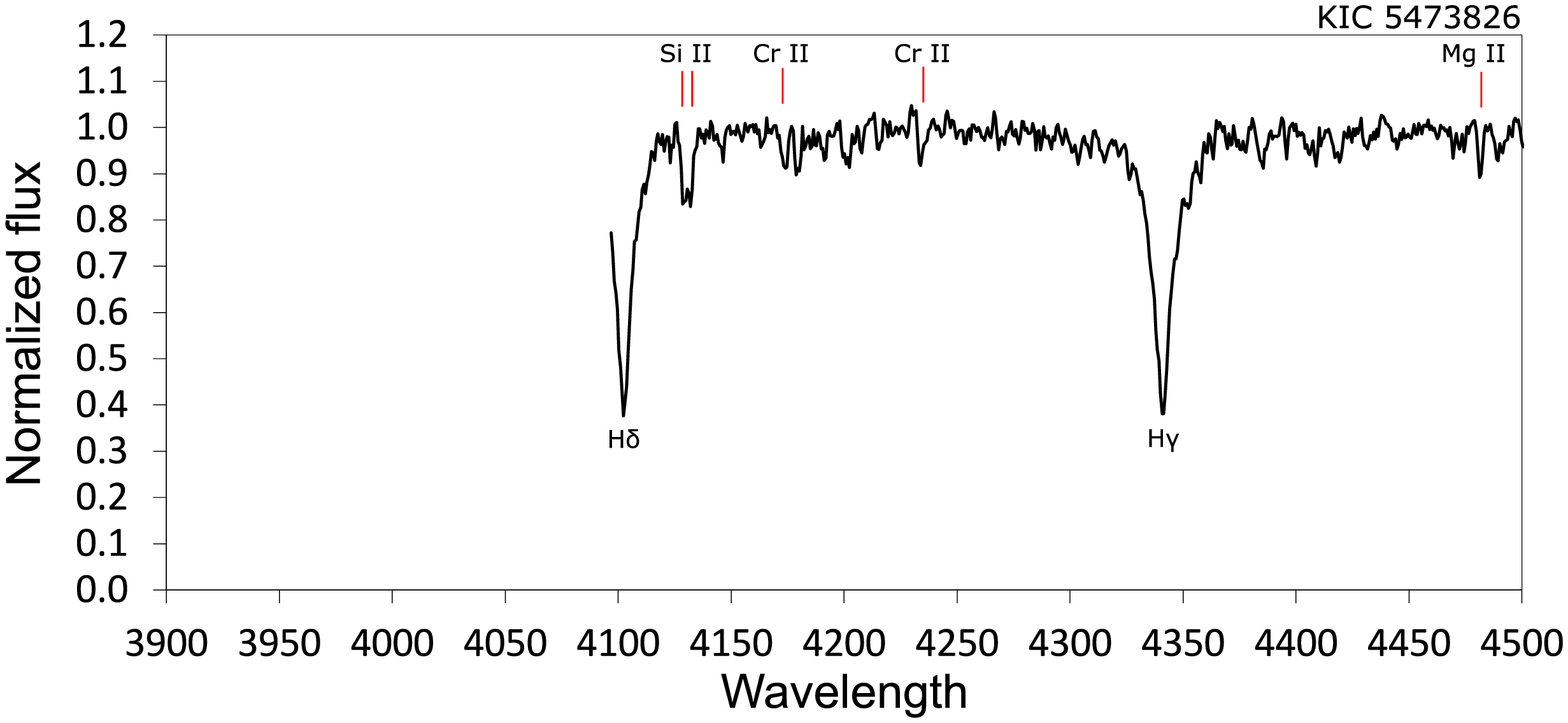}
        \includegraphics[width=\columnwidth]{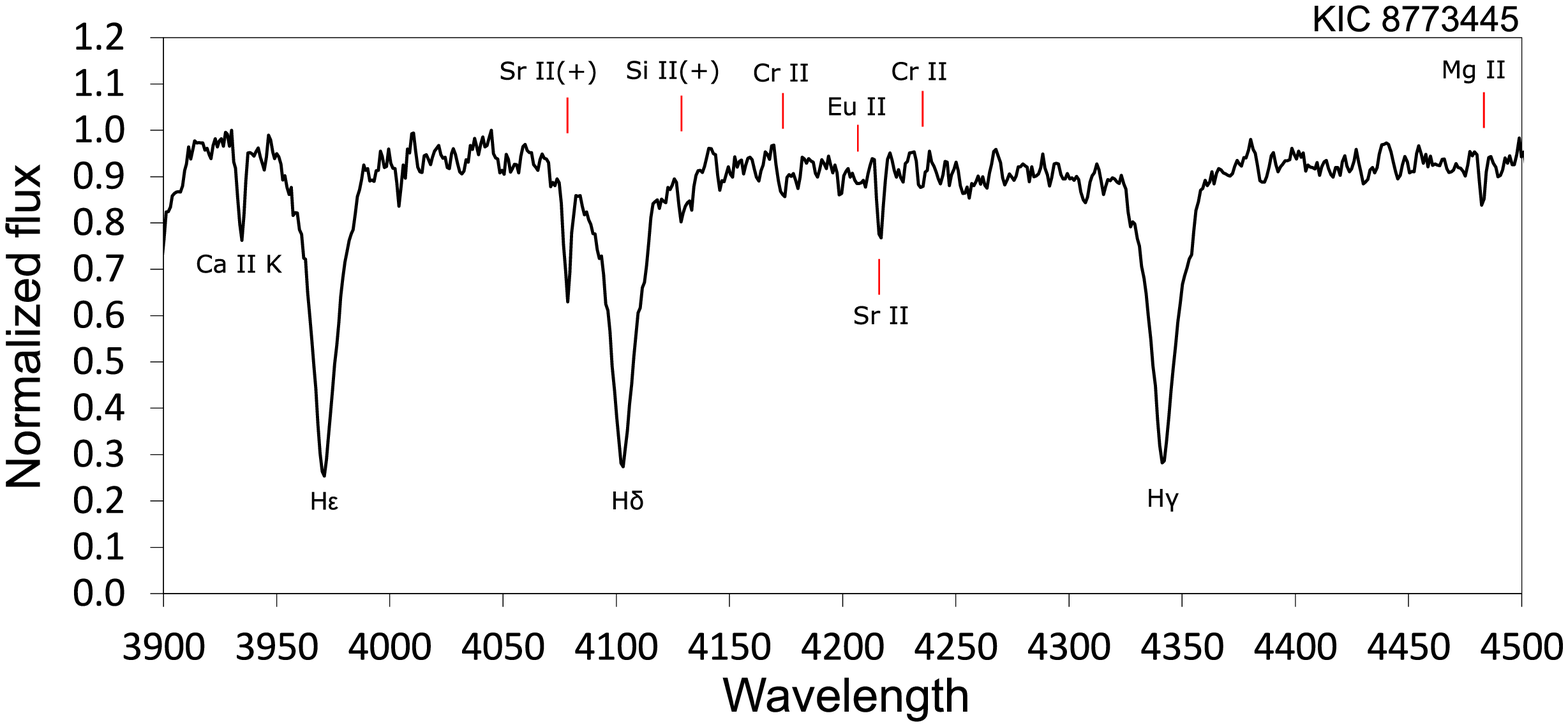}
        \includegraphics[width=\columnwidth]{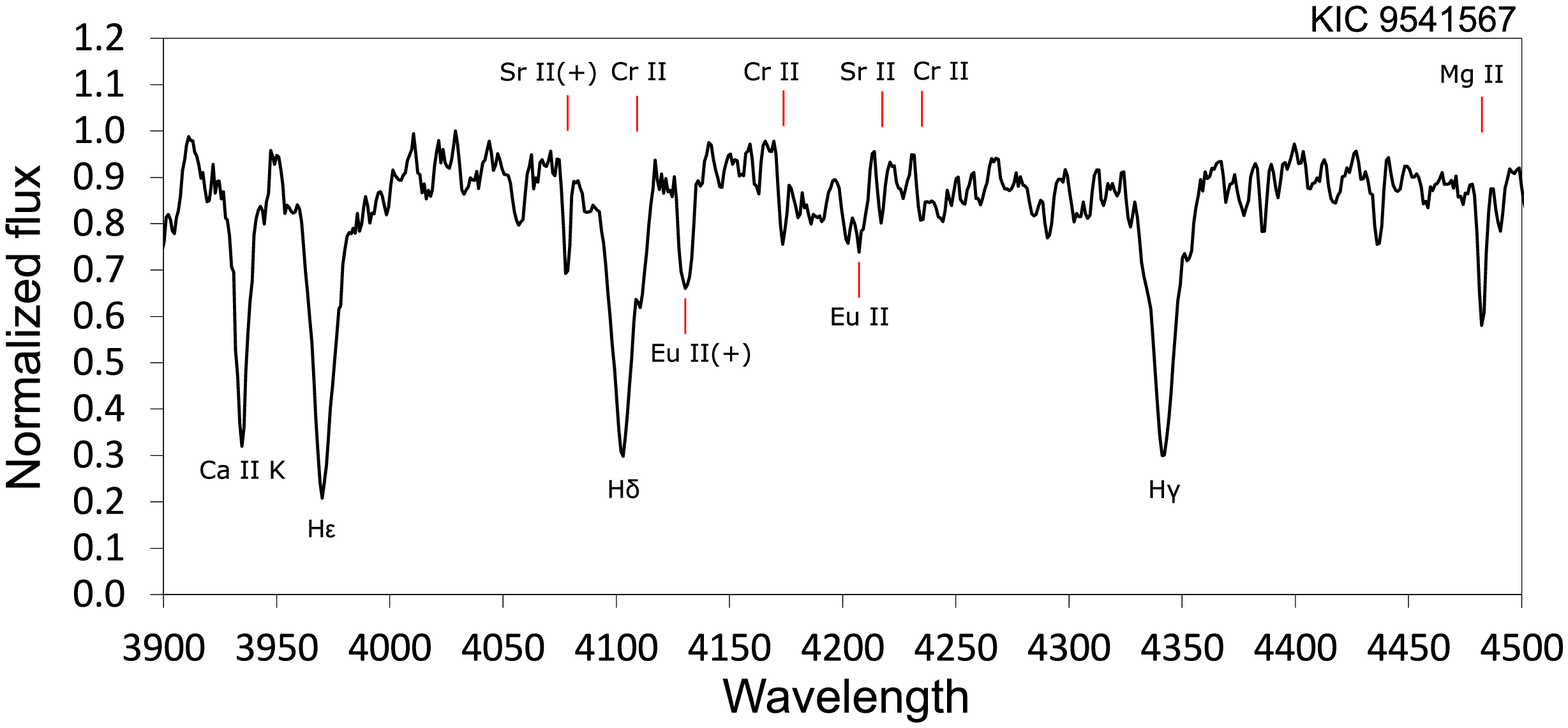}
    \caption{Blue-violet region of sample spectra that have been employed for spectral classification of our sample stars. The panels illustrate the SAO RAS spectrum of KIC\,5473826 (spectral type B9 V SiCr; upper panel) and the LAMOST DR2 spectra of KIC\,8773445 (spectral type A0 IV SiCrSrEu; middle panel) and KIC\,9541567 (spectral type A9 V SrCrEu; lower panel). Some lines of interest are indicated. Please note that at this resolution, the \ion{Ca}{H} line is blended with H$\epsilon$. The listed spectral types are from this investigation (cf. Table \ref{table_data1}).}
    \label{fig_spectra}
\end{figure}

We checked for the availability of \citet{gray16} classifications for our sample stars. As suggested by the aforementioned authors, only classifications based on spectra rated as "excellent" (average S/N\,=\,180), "very good" (average S/N\,=\,73), and "good" (average S/N\,=\,42) were considered. We found 32 matches and downloaded the corresponding LAMOST spectra. For some of our targets, more than one spectrum is available. In this case, we relied on the spectrum with the highest S/N ratio in the Sloan $g$ band, as indicated by the corresponding LAMOST catalogue. All spectra were manually (re)classified to check the accuracy of the MKCLASS code spectral types. Sample LAMOST spectra are provided in Fig. \ref{fig_spectra}.

%-----------------------------------------------------------------------

\subsection{Spectral classification} \label{classification}

\begin{figure}
        \includegraphics[width=\columnwidth]{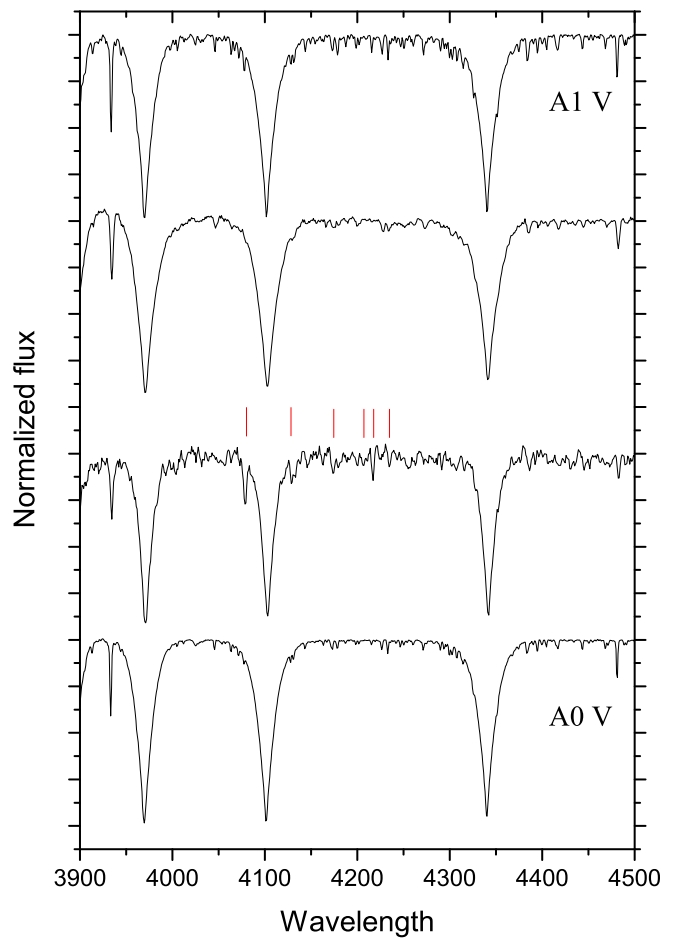}
    \caption{Comparison of the spectra of (from top to bottom) the MKK standard 48\,Cet (A1 V), the non-CP star KIC\,10550657 (A0 V), the mCP star KIC\,8773445 (A0 IV SiCrSrEu), and the MKK standard Vega (A0 V) in the blue-violet region. Some prominent lines of interest are indicated in the spectrum of KIC\,8773445 (for an identification of these lines, please refer to Fig. \ref{fig_spectra}). We also note the generally enhanced metal lines in this star, which are much stronger than in the A0 V standard despite an identical \vsini\ value.}
    \label{fig_spectra2}
\end{figure}

In this way, spectra were secured for 47 stars, including 46 photometric candidates and the CP2 star KIC\,8324268 = V2095\,Cyg. Spectral classification of the newly acquired spectra and the available LAMOST spectra was performed following current best practice and updated spectral standards. All programme stars were classified in the framework of a refined Morgan-Keenan-Kellman system (MKK hereafter), as described in \citet{gray87,gray89a,gray89b} and \citet{gray09}, using the standard techniques of MKK classification. For a precise classification of the objects and to identify possible peculiarities, the spectra were compared visually and overlaid with MKK standards. As an example, Fig. \ref{fig_spectra2} illustrates a comparison of the spectra of the non-CP star KIC\,10550657 (A0 V) and the mCP star KIC 8773445 (A0 IV SiCrSrEu) to the spectra of two MKK standards.

In this spectral range, the estimation of the effective temperature is based on the strength of the \ion{Ca}{K} line and the hydrogen lines. This, however, is not straightforward, as CP1 and CP2 stars may exhibit peculiar \ion{Ca}{K} profiles and line strengths \citep{faraggiana87,gray09}. Although attention was paid to this phenomenon, we caution that it may still have influenced the derived temperature types. We estimate the corresponding error to be $\pm$1\,subclass \citep{paunzen01}.

We emphasize that there is no modern (i.e. based on digital CCD spectra) classification scheme for CP stars available. CP stars are traditionally categorized into their respective subclasses (CP1/2/3/4) by comparing their spectra to MKK standards of the same effective temperature and searching for enhanced lines of certain elements. We adhered to this technique and also used the spectra of well-established CP stars to guide us in the identification of the various peculiarities. In general, the detection of spectral peculiarities has been straightforward with the achieved S/N ratio of the spectra.

Within the classification process, it is also important to consider rotational velocity. Since CP stars are generally slow rotators, MKK standards with a comparably low \vsini\ value need to be chosen for comparison; otherwise, errors in the luminosity classification might occur. (Sub)giants rotate much slower than main-sequence stars of the same temperature; therefore, the shape of the hydrogen lines of a CP\ star may provide a better match to an evolved star than a main-sequence object. This problematic situation has already been recognized using photographic prism spectra \citep{bidelman67} but is also an issue in automatic classification procedures such as MKCLASS. It might explain why many of our targets have been assigned a subgiant, giant, or as in the case of KIC\,8161798, even bright giant luminosity class by the MKCLASS code (Table \ref{table_data1}). We compared the spectra of all our targets with both high and low \vsini\ MKK standards. We note that \citet{paunzen99} has shown that luminosity class IV, which is commonly used in the MKK classification scheme, is not unambiguously defined; however, it is retained in this work for the classification process.

mCP stars exhibit several characteristic and unique flux depressions \citep{kupka03}. The most conspicuous depression is situated at around 5200\,\AA\ and is caused by contributions of Si, Cr, and Fe in the case of strongly overabundant surface metallicities and the presence of an organized magnetic field \citep{khan07}. We searched the available spectra for the presence of this feature and, if present, assessed its strength. The results are also given in Table \ref{table_data1}.

Except for the luminosity class, our derived spectral types are consistent and in good agreement with the classifications of \citet{gray16}; our classifications tend toward slightly earlier types for several stars. Most importantly, we confirmed the presence of peculiarities in all stars that have been identified as chemically peculiar by the MKCLASS code, although the derived peculiarity types differ in several objects. For example, the MKCLASS code identified enhanced Si lines in the star KIC 8773445, which was classified as A0 IV Si \citep{gray16}. We confirmed the presence of strong Si lines, but also identified enhanced lines of Cr, Sr, and Eu in the same LAMOST spectrum (see Fig. \ref{fig_spectra}). Consequently, the star was classified as A0 IV SiCrSrEu in the present study.

We identified four objects (KIC\,2441702, KIC\,3945892, KIC\,5814635, and KIC\,7628336) whose spectra are characterized by a general enhancement of the metal lines and thus bear strong similarities to the CP1 stars. Two of these stars (KIC\,3945892 and KIC\,5814635), however, show a mild 5200\,\AA\ flux depression, which is normally not found in CP1 stars \citep{kupka04}. Small changes in light curve shape are present in KIC\,5814635 and KIC\,7628336, which are not expected in mCP stars (cf. Sects. \ref{sample_selection} and \ref{periodogram}), but they are only present in parts of the data and likely of instrumental origin. Nevertheless, we caution that a detailed abundance analysis is necessary to shed more light on the true nature of these objects. We do not list separate spectral types (hydrogen-line type, K-line type, and metallic-line type) for these stars but  indicated these types by adding the suffix "metals strong" to the corresponding spectral types.

The derived spectral types are listed in the presentation of results in Table \ref{table_data1}. No peculiarities were found in the spectra of seven stars. Together with other objects of interest, these stars are discussed in Sect. \ref{nonCPs}.

%-----------------------------------------------------------------------

\subsection{Astrophysical parameters and Kiel diagram} \label{Kiel_diagram}

For the construction of a logarithmic surface gravity (\logg) versus logarithmic effective temperature (log\,\teff) diagram (also referred to as a Kiel diagram), we used the improved astrophysical parameters of \kepler\ targets published by \citet{mathur17}. As a first step, these parameters were compared to other sources.

To check the reddening estimates, several published reddening maps were employed for the stars with available parallaxes \citep{arenou92,chen98,schlegel98,drimmel03,green15}. In general, the reddening values derived from the maps are lower than those of \citet{mathur17} by a factor of three. This value is suspicious because it nearly exactly corresponds to the conversion value $R_V$ between $A_V$ and $E(B-V)$. However, we found no hint of wrongly listed parameters in any of the employed references and are currently unable to explain the observed discrepancies.

\begin{figure}
        \includegraphics[width=\columnwidth]{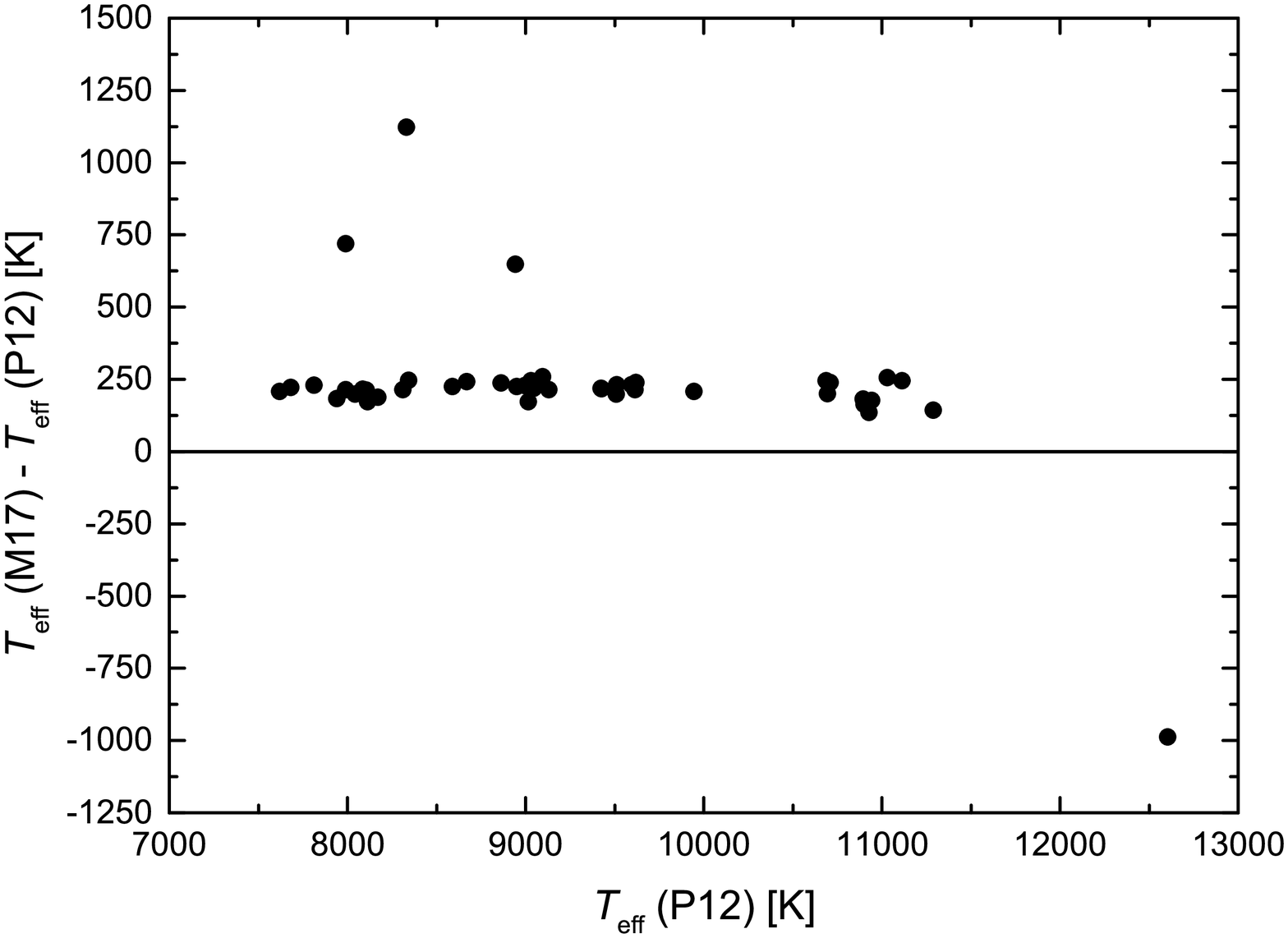}
    \caption{Comparison of the \teff\ values of \citet{pinsonneault12} (P12) and \citet{mathur17} (M17) for our sample stars.}
    \label{fig_temperature_scales}
\end{figure}

Unfortunately, Str{\"o}mgren-Crawford $uvby\beta$ photometry is not available for any of our target stars \citep{paunzen15b}, which prevents us from verifying the astrophysical parameters listed in \citet{mathur17}. We therefore compared the \citet{mathur17} \teff\ values with those from \citet{pinsonneault12}, who used Sloan Digital Sky Survey (SDSS) $griz$ filters to tie the fundamental temperature scale with correction terms for surface gravity effects, metallicity, and statistical corrections for binary companions or blending. With the exception of four stars, all 42 objects lie on a line with an offset of 215(29)\,K, in the sense that the \teff\ values of \citet{mathur17} are hotter by 215(29)\,K (cf. Fig. \ref{fig_temperature_scales}). There obviously is a statistically significant offset between the temperature scales, which is not caused by the CP nature of our sample stars.

To further investigate this matter, we derived \teff\ values of our target stars by employing the empirical calibration of \citet{paunzen17}, which is based on $(VJHK_{\rm s})$ photometry. This calibration is tied to the Str{\"o}mgren-Crawford $uvby\beta$ photometric system and does not depend on metallicity or rotational velocity. First, the $A_V$ values of \citet{mathur17} were employed. We find that in the range from 7\,000\,K to 10\,000\,K, the photometric calibrations yield hotter temperatures by about 1\,500\,K. Above this range, even hotter \teff\ values up to 20\,000\,K were derived, which correspond to a spectral type of about B2\,V. This is in striking contrast to the derived spectral types (Table \ref{table_data1}), as our sample does not contain any stars earlier (and hence hotter) than spectral type B7\,V. The only exception to this is the He-weak star KIC\,4136285 (spectral type B5 He wk; \citealt{RM09}). Temperature-wise, the upper limit for a B8\,V star is about 14\,000\,K \citep{pecault13}.

However, if we employ the $A_V$ values of the reddening maps, which are a factor three less, the temperatures in the range from 7\,000\,K to 10\,000\,K show a band width of about $\pm$700\,K around a perfect linear relationship. At hotter temperatures, the behaviour becomes ambiguous. The photometric calibration yields effective temperatures up to 18\,000\,K, which is still far too hot for the investigated spectral types. It is well known that CP2 stars show a "blueing" effect \citep{maitzen80}: owing to stronger UV absorption than in chemically normal stars, photometric calibrations yield hotter temperatures, which might partly explain the observed discrepancies. Nevertheless, in summary, the situation of the \teff\ values is not satisfactory and can only be improved by high-resolution spectroscopy or intermediate-band photometry.

An independent verification of the \logg\ values listed by \citet{mathur17} is very difficult. Broadband $(VJHK_{\rm s})$ photometry does not allow us to calibrate this parameter. In order to check the absolute magnitude estimates, we calculated absolute magnitudes for the eight stars that boast parallaxes with errors less than 25\% in the Gaia TGAS \citep{Gaia1}. The derived values match those from \citet{mathur17} within the errors, although the uncertainties are large.

\begin{figure}
        \includegraphics[width=\columnwidth]{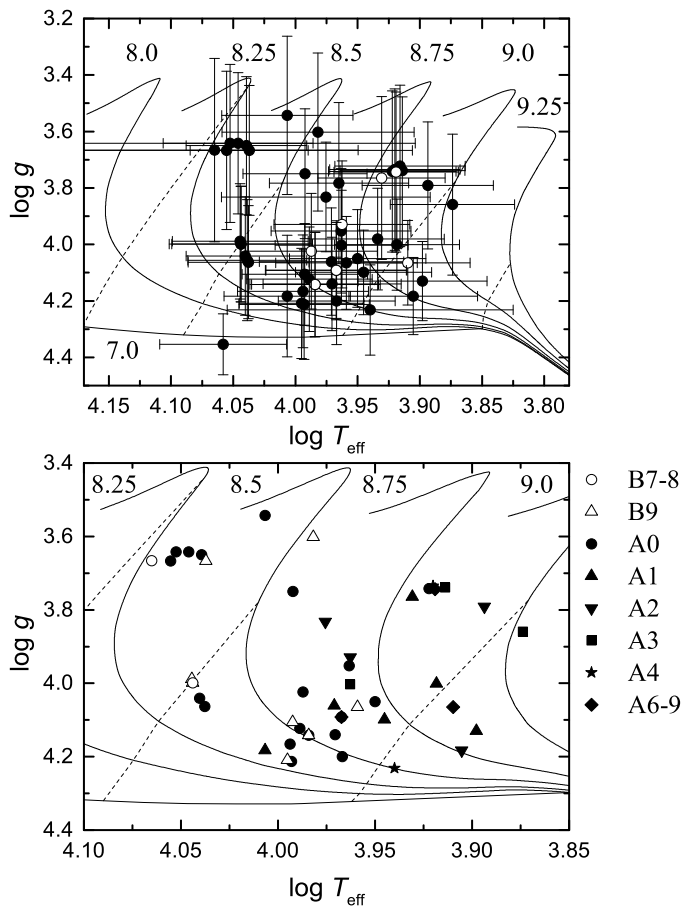}
            \caption{Kiel diagram indicating the location of our sample stars and the corresponding errors. The solid lines denote PARSEC isochrones \citep{bressan12} for solar metallicity ($[Z]$\,=\,0.019\,dex) and different ages from log\,$t$\,=\,7.0 to 9.25, as indicated. The dotted lines give the positions for different masses of $M_{\odot}$\,=\,1.5, 2, 3, and 4, respectively. The seven stars that have not been found to exhibit chemical peculiarities in this study are indicated by open symbols in the upper panel. The lower panel shows a zoomed-in version that includes information on the spectroscopic classifications, as indicated in the legend. We note that for the CP4 star KIC\,4136285, no parameters are available in the \citet{mathur17} catalogue and the star is therefore missing in the diagram.}
    \label{fig_HRD}
\end{figure}

The above listed limitations have to be kept in mind when considering the resulting Kiel diagrams, which are shown in Fig. \ref{fig_HRD}. Also given are PARSEC isochrones \citep{bressan12} for solar metallicity ($[Z]$\,=\,0.019\,dex) and different ages from log\,$t$\,=\,7.0 to 9.25, as indicated. The dotted lines indicate the positions for different masses of $M_{\odot}$\,=\,1.5, 2, 3, and 4, respectively. The seven stars that were not found to exhibit chemical peculiarities in this study are indicated by open symbols in the upper panel of Fig. \ref{fig_HRD}. We note that for the CP4 star KIC\,4136285, no parameters are available in the \citet{mathur17} catalogue, and the star is therefore missing in the diagram.

From this figure, we conclude that our targets populate the whole main sequence or luminosity classes V and IV, i.e. no giants are among our sample stars. In addition, no stars lie below the zero-age main sequence (ZAMS), which gives us confidence in the \logg\ values from \citet{mathur17}. This result is in line with our knowledge about the evolutionary status of CP stars \citep{netopil17}. The observed mass interval ranges from 1.5 $M_{\odot}$ to 4\,$M_{\odot}$, which is due to the applied target selection criteria (cf. Sect. \ref{target}). The derived errors on the parameters are large and preclude estimating the individual masses to better than $\pm$1\,$M_{\odot}$, which is not sufficient for the calibration of precise masses and ages. Therefore, a further statistical investigation of possible correlations between the rotation period and other astrophysical parameters is not possible.

The lower panel of Fig.\ref{fig_HRD} gives a detailed view and indicates the distribution of spectral types among our sample stars. Most of our stars have a spectral type of A0\,V (Table \ref{table_data1}). The \teff\ values of the A0\,V stars range from 8\,350\,K to 11\,350\,K (mean of 9\,980\,K and median of 9\,820\,K). For reference A0 dwarf stars, \citet{gerbaldi99} derived a \teff\ range from 9\,000\,K to 11\,000\,K and almost identical mean and median values, which supports the validity of these parameters. A comparable spread is seen for the other spectral types. The derived spectral types are therefore in good agreement with the \teff\ values of \citet{mathur17}. We conclude that the homogeneously derived astrophysical parameters of \citet{mathur17} are useful in a statistical sense.

%-----------------------------------------------------------------------

\subsection{\kepler\ data and linear period analysis} \label{dataan}

We downloaded the \kepler\ PDC flux data for all target stars from the Mikulski Archive for Space Telescopes (MAST).\footnote{https://archive.stsci.edu/kepler/} The individual data sets were inspected and separately treated by several dedicated codes, predominantly our own, that respect the nature of the studied objects and the properties of the employed data.

\subsubsection{Periodograms, amplitude spectrograms, and light curve development} \label{periodogram}

For a first investigation of the character of the light variations, we used standard periodograms and mainly own amplitude spectrograms, which enable the estimation of the variability amplitude in any given frequency \citep{paunzen13,mikulasek15}, as well as other standard procedures (PERIOD04, \citealt{period04}; \citealt{press89}). We note that the results from the various methods are in excellent agreement.

As a rule, the periodogram codes were applied on original, unadjusted \kepler\ data to select targets with frequency spectra indicative of a single independent frequency and corresponding harmonics, as is typically observed in ACV variables (cf. Sect. \ref{target}). As a next step, we investigated the stability of the light curve shape. As ACV variables are found to exhibit stable light curves during long periods of time (cf. Sect. \ref{introduction}), no significant changes in light curve shape are expected during the four years of \kepler\ coverage. To this end, a principal component analysis (PCA) code termed DEVELOPMENT was applied on the unadjusted data. Changes in the amplitude of the higher principle components may indicate changes in light curve shape or variation of the period.

\begin{figure}
\begin{center}
        \includegraphics[width=0.9\columnwidth]{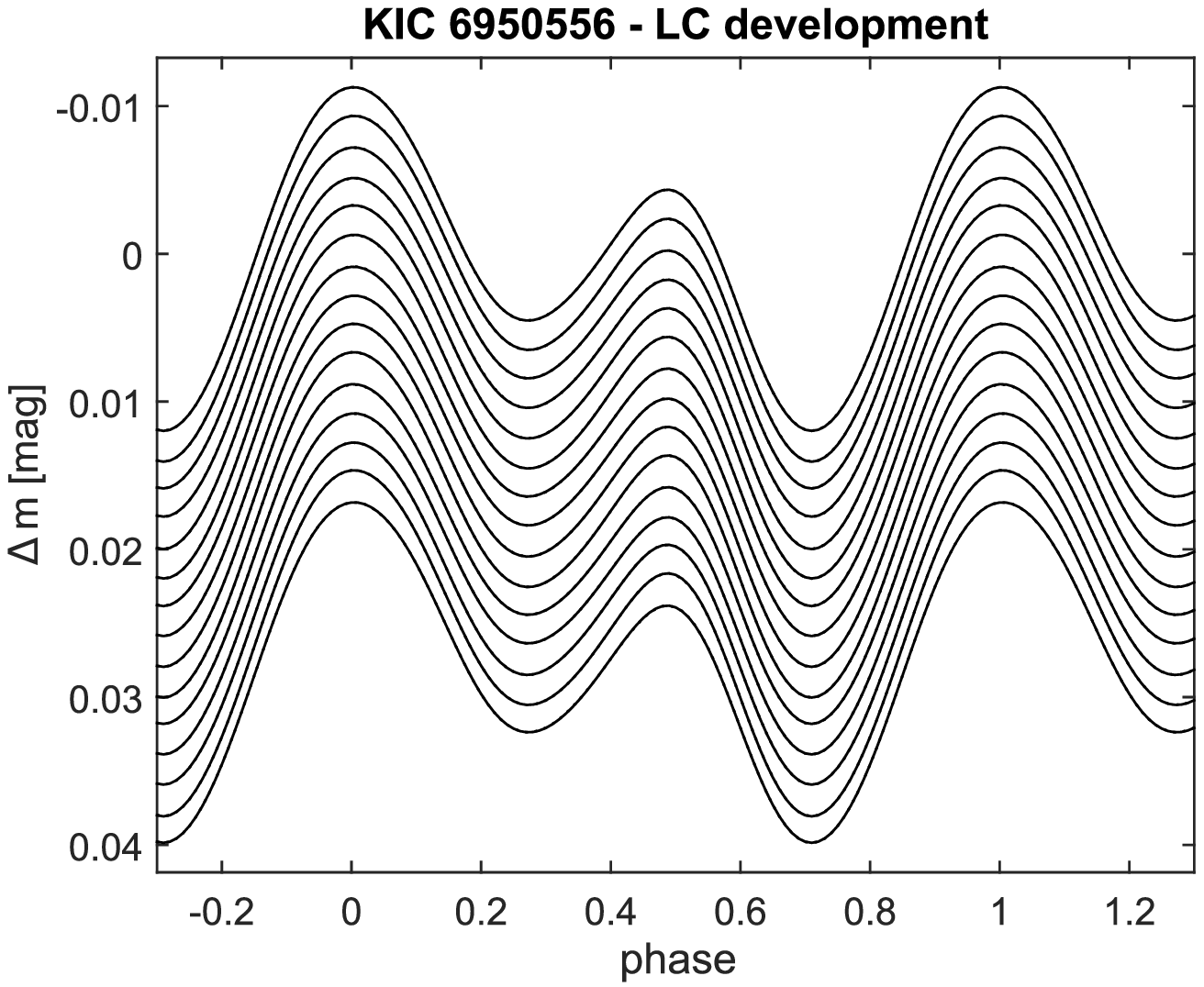}
    \includegraphics[width=0.9\columnwidth]{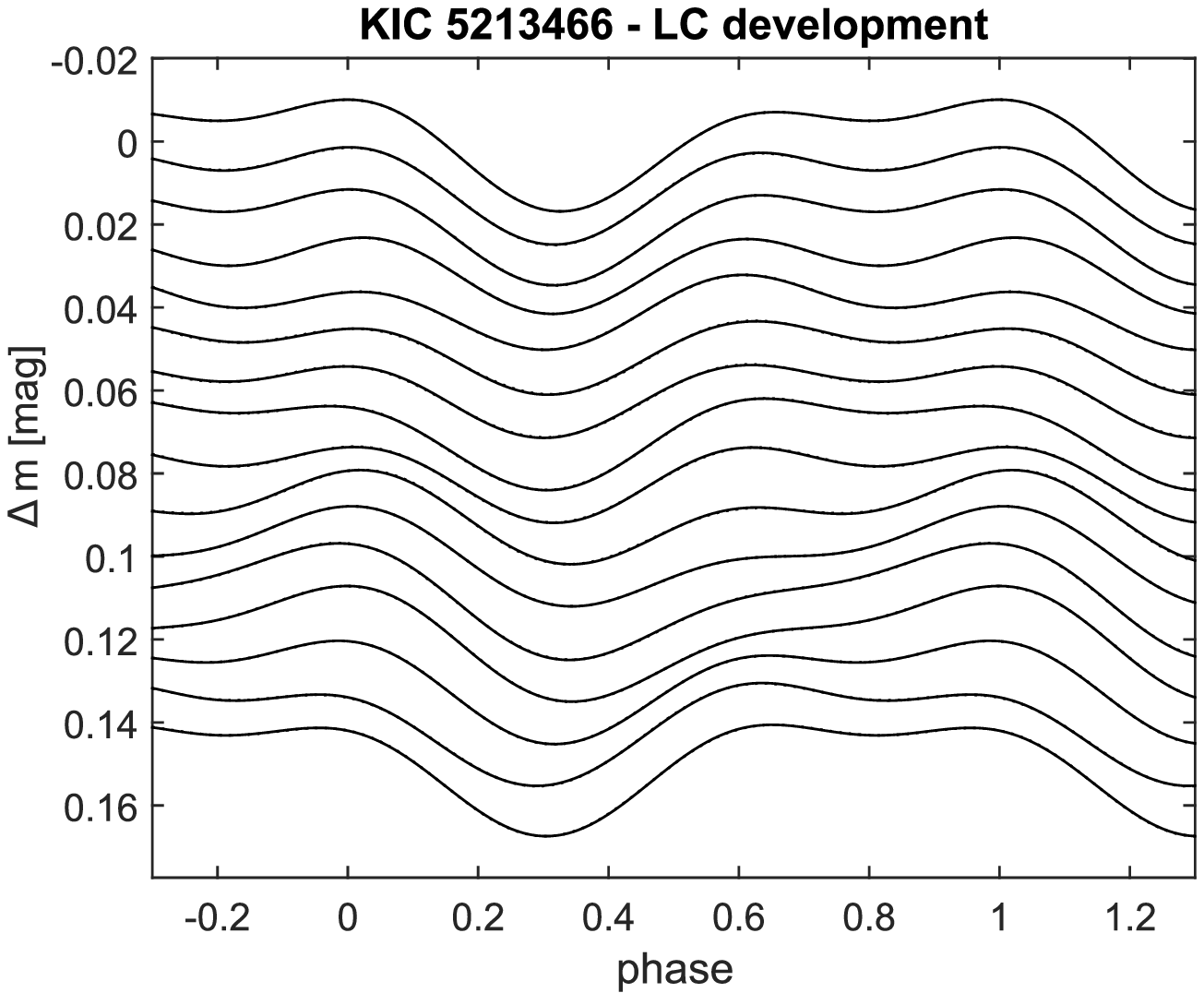}
    \caption{Light curve development of KIC\,6950556 (A0 V Si; upper panel) and KIC\,5213466 (A1 V; lower panel). The plots have been created with the DEVELOPMENT code and based on original, non-detrended Kepler PDC flux. The time distance between subsequent mean light curves is a quarter of a year.}
    \label{development}
\end{center}
\end{figure}

Examples of the output of the DEVELOPMENT code are presented in Fig. \ref{development}, which illustrates the results for KIC\,6950556, an ACV variable, and KIC\,5213466, a chemically normal object. No changes are apparent in the mean light curve of KIC\,6950556 (spectral type A0 V Si), which exhibits a stable light curve shape throughout the time period covered by \kepler. On the other hand, the star KIC\,5213466 (spectral type A1 V) shows dramatic changes in its mean light curve, which we cannot explain by instrumental effects. Together with the non-detection of peculiarities in the spectrum, this disqualifies KIC\,5213466 as a member of the class of ACV variables.

\subsubsection{Linear ephemerides and light curves from adjusted data} \label{ephemerides}

A detailed inspection of short segments of the light curves highlights the impressive accuracy of the \kepler\ photometry, which achieves a precision of the order of some tenths of a micromagnitude. \kepler\ data, therefore, should be well suited to the goals of the present investigation.

Unfortunately, despite careful processing in the pipeline reductions (cf. Sect. \ref{kepler}) and the removal of apparent outliers, \kepler\ photometric data are still significantly affected by instrumental effects. These occur mostly on longer timescales and diminish the quality of the data in so much as the actual accuracy may be lower than the theoretical accuracy limit by a factor of 30 for a given object. Although we were not able to remove these instrumental effects
entirely, their impact can be considerably reduced by the assumption that the light curve and period of an mCP star are constant and devoid of sudden excursions. This assumption is in accordance with our current understanding of mCP stars (cf. Sect. \ref{target}) and the results of this study. By modelling the expected light curve shape, we can extract the intrinsic light variations and study the residuals, which are dominated by instrumental effects and blurred by true (nearly-Gaussian) scatter.

\begin{figure}
\begin{center}
    \includegraphics[width=\columnwidth]{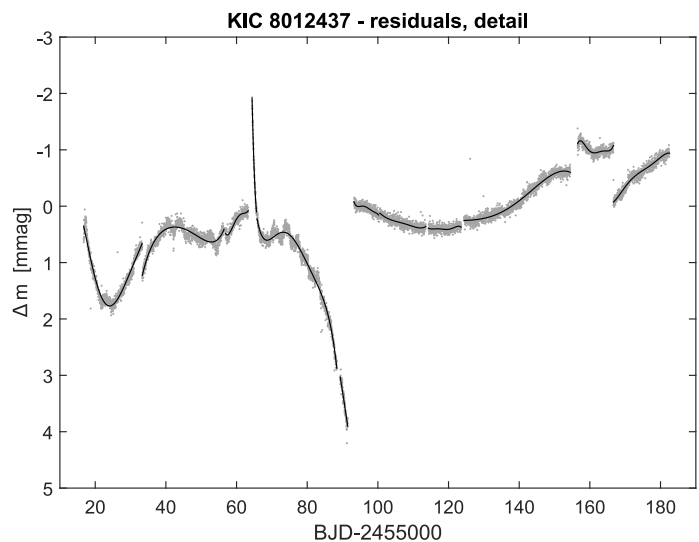}
    \caption{Dependence of the individual residuals (grey dots) on time in days, using the example of KIC\,8012437 (spectral type A4 V SiCr). The full lines denote Chebyshev polynomials up to the sixth degree that were fitted to the individual segments.}
    \label{residuals}
\end{center}
\end{figure}

An examination of the residuals in several tens of periodically variable or invariable \kepler\ stars shows that they are clustered alongside more or less horizontal segments, separated, as a rule, by gaps. These trends can be well fitted by Chebyshev polynomials up to the sixth order (see Fig. \ref{residuals}). The coefficients of the fits were found using our own implementation of the robust regression that eliminates remnant outliers \citep{mikulasek03}. After correction for these trends, we obtain a detrended, very accurate light curve with an uncertainty of about 0.1\,mmag.

During some parts of the \kepler\ light curves, changes in the amplitude of the photometric variations are apparent, which may escalate the uncertainty of the results to the point of being useless. Short-term changes in amplitude are not expected in mCP stars, which are known to exhibit stable spot configurations over long periods of time, and, suspiciously, the observed amplitude variations seem to affect separate quarters of \kepler\ data differently. These variations are surely of instrumental origin and we attribute them to transient changes in the sensitivity of the photometric apparatus. Assuming a constant or predictably varying light curve, this situation can be improved by a careful rescaling of the corresponding data intervals. However, instrumental oscillation reaching 1\,mmag, which we have not been able to eliminate, still remain in several data segments. Nevertheless, by a careful detrending and rescaling of \kepler\ data, a gain in accuracy by a factor of about ten could be achieved for most objects.

Employing the outlined methodology, we derived linear ephemeris parameters, effective amplitudes, and standard deviations of the light curve fit, which are presented below (Table \ref{table_data1}). The effective amplitude $A_{\rm{eff}}$ is a robust measure of the extent of the variability of periodically variable objects and was introduced by \citet{mikulasek07}, who defined it as follows:

%\begin{adjustbox}{max width=\textwidth}

\begin{equation}
A_{\rm{eff}}^2 = 8\int_0^1{F^2 (\varphi) \mathrm{d} \varphi},\quad \mathrm{if}
\end{equation}
\begin{equation}
F(\varphi) =  \!\sum_{i=1}^k a_i \cos (2 \pi i \varphi )\!+\!b_i \sin (2\pi i \varphi), \ A_{\mathrm{eff}}^2=4\!\sum_{i=1}^k a_i^2\!+\!b_i^2,
\end{equation}

%\end{adjustbox}

where $\varphi$ is the common phase and $F(\varphi)$ a periodical function that represents the light curve. The \kepler\ light curves of our target stars can be expressed by a harmonic polynomial of the $k-$th order (typically: $k\sim12$). $\{a_i,b_i\}$ are coefficients of the Fourier decomposition. In the case that $k=1$, the effective amplitude equals the common amplitude. The detrended and rescaled \kepler\ data, which are corrected for the possible trends identified above, are employed for the construction of the phase plots provided in Fig. \ref{allphased}.

%-----------------------------------------------------------------------

\section{Results} \label{result}

Employing the methodology outlined in Sect. \ref{dataso}, we identified 51 photometric ACV candidates by investigating the \kepler\ light curves of early-type stars. In addition, two well-confirmed mCP stars (KIC 4136285 and KIC 8324268) were included into the sample. Using newly acquired and archival spectra, we were able to investigate 46 photometric candidates, 39 of which (85\%) turned out to be bona fide mCP stars. This result indicates that our photometric selection criteria constitute a viable and efficient method of identifying mCP star candidates in massive photometric time series databases. 

Five photometric candidates remain either without spectrum (KIC\,2969628, KIC\,3326428, KIC\,6278403, KIC\,8362546) or recent spectroscopic observations (KIC\,11560273). These stars are presented as candidate ACV variables, which should be confirmed by spectroscopy. For seven stars, no evidence for chemical peculiarities has been found in this investigation and \citet{gray16} (KIC\,5213466, KIC\,5727964, KIC\,8415109, KIC\,8569986, KIC\,10082844, KIC\,11671226) or our own spectra (KIC\,10550657).

In summary, the final sample consists of 41 confirmed ACV variables, 5 candidate ACV variables, and 7 stars in which no chemical peculiarities could be established. The latter objects are potentially of great interest and are discussed in more detail in Sect. \ref{nonCPs}. During our literature search, we noted that several of our photometric candidates have been entered as candidate ACV variables into the International Variable Star Index (VSX; \citealt{VSX}) of the American Association of Variable Star Observers (AAVSO) by observer R. Jansen.

Table \ref{table_data1} presents essential data for our sample stars and is organized as follows:

\begin{itemize}
\item Column 1: KIC number.
\item Column 2: alternative identifier (e.g. HD number).
\item Column 3: right ascension (J2000). Positional information was taken from Gaia DR1 \citep{Gaia1,Gaia2} and, in some cases, from the UCAC4 catalogue \citep{UCAC4}.
\item Column 4: declination (J2000).
\item Column 5: spectral type from the literature. If not indicated otherwise, all literature spectral types were extracted from \citet{gray16}. For some stars, more than one classification is available in this source; in these cases, the classification corresponding to the highest Sloan $g$ band S/N value was chosen. Quality flags and the $g$ band S/N value are provided in parentheses. Spectral classifications from other sources are highlighted in italics and the corresponding reference is given.
\item Column 6: spectral type, this paper.
\item Column 7: flux depression at around 5200\,\AA\ (N=no depression, M=mild depression, S=strong depression).
\item Column 8: period (d), given to the last significant digit, and corresponding uncertainties.
\item Column 9: epoch (M0-2400000 [BJD]); time of maximum is indicated.
\item Column 10: effective amplitude, $A\textsubscript{eff}$[mmag], following \citet{mikulasek07}.
\item Column 11: scatter, $s$[mmag].
\item Column 12: number of observations available, $N$.
\item Column 13: number of observations included in the final analysis, $N\textsubscript{eff}$.
\item Column 14: number of rejected observations, $N\textsubscript{k}$.
\item Column 15: estimated magnitude in the Kepler band ($Kp$) from the Kepler Input Catalogue (KIC; \citealt{KIC}).
\end{itemize}

The astrophysical parameters of our target stars, which have been taken from \citet{mathur17} and employed in the construction of the Kiel diagram shown in Fig. \ref{fig_HRD}, are given in Table \ref{table_data2}, which is organized as follows:

\begin{itemize}
\item Column 1: KIC number.
\item Column 2: effective temperature, \teff.
\item Column 3: upper 1$\sigma$ confidence interval in \teff\ (\teff\ err$+$).
\item Column 4: lower 1$\sigma$ confidence interval in \teff\ (\teff\ err$-$).
\item Column 5: logarithmic surface gravity, \logg.
\item Column 6: upper 1$\sigma$ confidence interval in \logg\ (\logg\ err$+$).
\item Column 7: lower 1$\sigma$ confidence interval in \logg\ (\logg\ err$-$).
\end{itemize}

The light curves of all objects, based on detrended and rescaled \kepler\ data (cf. Sect. \ref{ephemerides}) and phased with the linear ephemerides presented in Table \ref{table_data1}, are illustrated in Fig. \ref{allphased}.

\begin{sidewaystable*}
\caption{Essential data for the 53 confirmed or candidate mCP stars identified in the present paper, sorted by KIC number. Columns are explained in the text (Sect. \ref{result}).}
\label{table_data1}
\scriptsize{
\begin{center}
\begin{adjustbox}{max width=\textwidth}
\begin{tabular}{lllllllllllllll}
\hline
\hline 
KIC       &     altID   &       RA          &   Dec         &   SpT       &      SpT          & dep.           & P       &       M0-2400000      &       Aeff   & s      &      N     &        Neff   &       Nk      &       Mag      \\
      &       & [J2000] & [J2000] & [lit] & [this paper] & $\lambda$5200\AA & [d]  & [BJD]        & [mmag] & [mmag] &       &         &     & [Kp] \\
\hline
1576144 &       TYC 2666-407-1  &       19 27 55.197    &       +37 10 32.18    &       A3 V SrSi (vgood, 57)      &       A3 V SiSrEu     &       M       &       2.68671461(24)         &       55679.7051      &       23.8    &       0.18    &       65264   &       64635   &       126     &       11.251  \\
1865567 &       2MASS J19231341+3722243 &       19 23 13.411    &       +37 22 24.33        &       B8 IV-V Si (vgood, 319) &       B9 V Si &       M       &       1.34943589(8)         &       55679.8336      &       4.9     &       0.05    &       65243   &       64997   &       113     &       10.569  \\
2441702 &       TYC 3134-1731-1 &       19 24 16.228    &       +37 47 48.93    &       A1 III-IV SrSi (good, 247) &       A2 V SiSrCrEu metals strong     &       N       &       8.185116(14)         &       55683.6693      &       0.6     &       0.05    &       65264   &       65089   &       90      &       10.292  \\
2853320 &       2MASS J19263017+3802518 &       19 26 30.173    &       +38 02 51.73        &               &       A0 V Si &       S       &       5.0653265(22)         &       55681.1373      &       9.5     &       0.22    &       64795   &       64651   &       122     &       13.705  \\
2969628 &       TYC 3120-750-1  &       19 02 54.741    &       +38 09 57.48    &               &               &               &       1.97361145(14)         &       55681.6167      &       11.8    &       0.11    &       65263   &       65058   &       123     &       11.698  \\
3326428 &       2MASS J19061861+3824209 &       19 06 18.610    &       +38 24 20.77        &               &               &               &       7.7004207(15)         &       55679.8865      &       26.9    &       0.22    &       65261   &       65052   &       197     &       13.413  \\
3561656 &       TYC 3136-717-1  &       19 42 08.155    &       +38 41 37.88    &               &       A0 V Si    &       M       &       1.91715050(28)  &       55681.7191      &       3.9     &       0.10    &       52279   &       52162   &       73      &       11.665  \\
3945892 &       TYC 3121-127-1  &       19 16 08.953    &       +39 00 25.06    &       kA2hA7mA6 SiEu (n/a, 125) &       A2 V SiSrCrEu metals strong     &       M       &       4.0832255(6)         &       55679.6350      &       27.9    &       0.21    &       65256   &       64972   &       113     &       12.228  \\
4136285 &       V545 Lyr, HD 176582     &       18 59 12.292    &       +39 13 02.36        &       \textit{B5 He wk}$^{1}$ &               &               &       1.5819866(6)         &       55681.4879      &       23.2    &       0.08    &       7793    &       7705    &       24      &       6.510   \\
4828345 &       TYC 3125-287-1  &       19 18 25.729    &       +39 55 56.98    &       B8 IV Si (vgood, 64)       &       B8 V Si &       M       &       0.841079046(15) &       55680.6663      &       34.9    &       0.21    &       65230   &       64427   &       88      &       12.780  \\
5213466 &       2MASS J19523118+4023594 &       19 52 31.176    &       +40 23 59.37        &       A1 V (vgood, 42)        &       A1 V    &       N       &       2.819508(3)         &       55680.6120      &       20.3    &       2.50    &       65266   &       65049   &       81      &       13.075  \\
5473826 &       HD 226339       &       19 52 51.075    &       +40 36 21.49    &               &       B9 V SiCr  &       S       &       1.051203405(16) &       55681.4770      &       29.3    &       0.14    &       65264   &       63961   &       119     &       10.885  \\
5727964 &       2MASS J19501553+4058357 &       19 50 15.530    &       +40 58 35.63        &       A7 V (vgood, 34)$^{2}$  &       A6 V    &       N       &       1.63013818(8)         &       55680.7668      &       9.1     &       0.16    &       64797   &       64659   &       83      &       12.925  \\
5739204 &       2MASS J19591596+4056166 &       19 59 15.967    &       +40 56 16.54        &               &       B9 V SiEu       &       M       &       1.81123002(11)         &       55681.5302      &       32.5    &       0.27    &       65267   &       64986   &       89      &       13.608  \\
5774743 &       TYC 3124-443-1  &       19 04 20.283    &       +41 01 45.07    &       A6 V (Sr)SiEu (vgood, 118) &       A3 V SiCr       &       M       &       4.0735673(5)         &       55680.5594      &       12.1    &       0.11    &       50636   &       50483   &       81      &       12.136  \\
5814635 &       TYC 3141-42-1   &       19 51 20.384    &       +41 03 29.91    &               &       A1 V SiSrCrEu metals strong        &       M       &       3.8696063(5)    &       55686.1321      &       18.4    &       0.15    &       65266   &       64785   &       111     &       11.399  \\
6065699 &       HD 188101       &       19 51 50.998    &       +41 20 55.69    &               &       B7 V Si    &               &       3.9873152(4)    &       55681.4953      &       28.7    &       0.15    &       64790   &       64644   &       104     &       7.833   \\
6206125 &       TYC 3142-25-1   &       19 28 49.306    &       +41 35 50.21    &       A1 IV-V Si (vgood, 404)    &       B9 V (Si)       &       N       &       3.596345(6)         &       55682.2759      &       8.1     &       1.23    &       65267   &       65295   &       99      &       10.644  \\
6278403 &       HD 181436       &       19 18 36.092    &       +41 36 43.38    &               &               &               &       1.19123857(4)         &       55680.8187      &       7.2     &       0.05    &       65264   &       64603   &       94      &       8.758   \\
6305572 &       2MASS J19491098+4138090 &       19 49 10.975    &       +41 38 09.05        &               &       A0 V Si &               &       1.50011965(9)         &       55681.6993      &       22.9    &       0.24    &       64797   &       64569   &       103     &       13.825  \\
6310414 &       HD 226421       &       19 53 44.325    &       +41 41 04.26    &       B8 IV-V Si (vgood, 358)    &       B9 V Si &       S       &       9.602575(5)         &       55683.8553      &       13.6    &       0.06    &       65266   &       64580   &       209     &       10.239  \\
6426158 &       HD 177128       &       19 01 32.491    &       +41 51 59.33    &       A1 IV-V SrSiEu (vgood, 509)        &       A1 V SiSrCrEu   &       S       &       4.8656355(7)         &       55683.2593      &       5.7     &       0.04    &       50642   &       50198   &       95      &       9.153   \\
6715809 &       2MASS J19511198+4206264 &       19 51 11.985    &       +42 06 26.27        &       A1 IV-V SrSi (vgood, 58)        &       A1 V SiCrEu     &       M       &       4.1979343(11)         &       55679.4407      &       17.9    &       0.28    &       65257   &       64846   &       110     &       12.503  \\
6864569 &       TYC 3142-919-1  &       19 29 04.114    &       +42 19 16.65    &       B9 V Si (vgood, 626)       &       B9 V Si &       M       &       2.32517273(4)         &       55682.4726      &       27.2    &       0.06    &       65263   &       64123   &       119     &       10.052  \\
6950556 &       2MASS J19294376+4229306 &       19 29 43.768    &       +42 29 30.59        &       A8 mB4 V Lam Boo (n/a, 145)     &       A0 V Si &       S       &       1.51178505(4)         &       55687.5095      &       18.9    &       0.13    &       64783   &       64512   &       125     &       12.751  \\
7628336 &       TYC 3148-183-1  &       19 49 36.246    &       +43 13 08.28    &       kA3hA6mA9 SiEu (n/a, 108) &       A3 V SiSrCrEu metals strong     &       N       &       2.5388345(5)         &       55682.2343      &       6.1     &       0.13    &       65257   &       63983   &       107     &       11.348  \\
7778838 &       TYC 3149-1303-1 &       19 56 44.960    &       +43 29 51.37    &       B9 V Si (vgood, 319)       &       A0 V SiCr       &       S       &       5.8766553(10)         &       55684.8697      &       25.3    &       0.15    &       65266   &       64731   &       109     &       11.868  \\
8012437 &       TYC 3131-1074-1 &       18 58 21.295    &       +43 49 26.42    &       A6 V SrSi (vgood, 164)     &       A4 V SiCr       &       N       &       4.0750812(5)         &       55677.6865      &       16.8    &       0.12    &       65261   &       64003   &       111     &       11.519  \\
8161798 &       2MASS J19213515+4403022 &       19 21 35.150    &       +44 03 02.24        &       A0 II SrSiEu (vgood, 502)       &       A0 V SiCr       &       S       &       2.20296272(5)         &       55680.9555      &       90.5    &       0.41    &       65263   &       65196   &       112     &       10.471  \\
8324268 &       V2095 Cyg, HD 189160    &       19 56 50.155    &       +44 16 16.03        &       \textit{B8 Si}$^{3}$ &  A0 V SiCr       &               &       2.00912016(7)         &       55680.8367      &       27.1    &       0.11    &       64797   &       62332   &       120     &       7.955   \\
8362546 &       2MASS J19224722+4419143 &       19 22 47.238    &       +44 19 14.23        &               &               &               &       1.1081387(23)         &       55681.4749      &       2.0     &       0.91    &       64794   &       64154   &       97      &       15.694  \\
8415109 &       TYC 3131-1445-1 &       18 58 29.902    &       +44 29 34.99    &       A0 V (vgood, 202)  &       A0 V    &       N       &       0.93943830(8)   &       55681.1931      &       2.3     &       0.07    &       65260   &       64984   &       91      &       11.085  \\
8569986 &       2MASS J19420663+4438592 &       19 42 06.634    &       +44 38 59.20        &       A2 IV-V (vgood, 71)     &       A2 V    &       N       &       3.1331751(4)         &       55681.2501      &       18.1    &       0.20    &       65261   &       65022   &       108     &       13.427  \\
8619436 &       TYC 3133-70-1   &       19 19 20.178    &       +44 47 06.69    &       B8 IV-V Si (vgood, 216)    &       B9 V Si &       S       &       3.13319347(23)         &       55682.1813      &       21.0    &       0.12    &       64793   &       64258   &       105     &       11.806  \\
8773445 &       2MASS J19531426+4457124 &       19 53 14.260    &       +44 57 12.33        &       A0 IV Si (vgood, 75)    &       A0 IV SiCrSrEu  &       S       &       3.6607828(7)         &       55681.2271      &       18.7    &       0.25    &       64797   &       64649   &       109     &       13.841  \\
8881883 &       TYC 3543-971-1  &       19 21 51.354    &       +45 06 55.33    &       ?? (unclassifiable, 18), \textit{A2:}$^{4}$        &       A0 IV Si        &       S       &       3.32369473(28)         &       55684.8852      &       18.1    &       0.14    &       65267   &       64961   &       122     &       12.441  \\
9541567 &       TYC 3557-2111-1 &       19 49 37.017    &       +46 07 11.96    &       F0 V Sr (good, 65) &       A9 V SrCrEu     &       S       &       2.24569344(4)         &       55680.6277      &       44.6    &       0.10    &       52201   &       51848   &       74      &       11.872  \\
10082844        &       2MASS J19391258+4701085 &       19 39 12.576    &       +47 01 08.46        &       A0 V (vgood, 104)       &       A0 V    &       N       &       2.0833835(14)         &       55681.1226      &       25.2    &       2.70    &       64793   &       64741   &       66      &       13.693  \\
10281890        &       TYC 3560-3168-1 &       19 39 27.586    &       +47 18 26.45        &       B9.5 IV Si (good, 861)  &       A0V SiCrEu      &       S       &       1.9422501(15)         &       55676.2603      &       35.5    &       0.05    &       65247   &       63528   &       131     &       10.413  \\
10324412        &       HD 176436       &       18 58 02.386    &       +47 25 10.88        &               &       A0 V Si &               &       1.73149557(5)         &       55680.7844      &       21.0    &       0.11    &       65255   &       64819   &       145     &       8.248   \\
10550657        &       TYC 3561-1322-1 &       19 49 38.884    &       +47 42 11.72        &               &       A0 V    &       N       &       0.88491555(14)         &       55680.6169      &       22.0    &       1.20    &       52200   &       52080   &       98      &       12.240  \\
10685175        &       2MASS J19541717+4757502 &       19 54 17.184    &       +47 57 50.14        &       A6 IV (Sr)Eu (vgood, 91)        &       A4 V Eu &       N       &       3.1019943(4)         &       55682.3739      &       10.7    &       0.11    &       52201   &       52021   &       115     &       12.075  \\
10905824        &       2MASS J18544461+4820247 &       18 54 44.620    &       +48 20 24.83        &       A1 V SrSiEu (vgood, 142)        &       A1 V SiCr       &       S       &       2.71953544(25)         &       55681.2693      &       11.8    &       0.15    &       65261   &       65021   &       126     &       12.811  \\
10959320        &       2MASS J18481841+4828541 &       18 48 18.415    &       +48 28 53.88        &               &       A0 V SiCrSr     &       S       &       2.44557724(17)         &       55681.6376      &       18.4    &       0.21    &       65260   &       65005   &       133     &       13.190  \\
10971633        &       TYC 3547-2631-1 &       19 19 05.021    &       +48 27 52.50        &       A0 V (Sr)Si (vgood, 323)        &       A0 V SiSr       &       N       &       3.980710(3)         &       55679.5623      &       2.8     &       0.23    &       65259   &       65184   &       92      &       11.461  \\
10982373        &       TYC 3560-2861-1 &       19 39 36.713    &       +48 28 37.98        &       A3 IV-V SrSiEu (good, 76)       &       A2 V SiSrEu     &       M       &       2.00280473(10)         &       55680.1221      &       29.6    &       0.18    &       65264   &       64857   &       111     &       11.696  \\
11098975        &       TYC 3562-258-1  &       19 52 31.799    &       +48 40 38.40        &               &       A0 V SiSr       &       M       &       0.99511463(3)         &       55687.9624      &       13.0    &       0.13    &       52181   &       51787   &       118     &       12.249  \\
11154043        &       TYC 3566-325-1  &       19 54 35.240    &       +48 47 04.64        &               &       A0 V SiCr       &       M       &       4.5298431(8)         &       55635.8824      &       12.3    &       0.13    &       51732   &       51525   &       116     &       11.991  \\
11465134        &       TYC 3565-508-1  &       19 45 38.824    &       +49 22 27.89        &               &       A0 V Si &       M       &       1.48780697(7)         &       55681.8720      &       17.5    &       0.13    &       65251   &       64964   &       127     &       12.399  \\
11560273        &       HD 184007A      &       19 29 55.297    &       +49 30 31.72        &       \textit{A0V, NErn of 2" cpm pair}$^{5}$ &               &               &       1.82746769(9)         &       55686.3430      &       2.8     &       0.03    &       64783   &       64380   &       133     &       8.036   \\
11671226        &       TYC 3565-655-1  &       19 45 27.840    &       +49 46 30.77        &       A6 IV-V (vgood, 320)    &       A5 V    &       N       &       2.9571620(3)         &       55681.4376      &       7.5     &       0.07    &       65264   &       64989   &       145     &       10.977  \\
11807603        &       TYC 3550-977-1  &       19 13 05.806    &       +50 00 13.38        &       A4 IV SrSiEu (vgood, 223)       &       A1 V SiSr       &       N       &       1.57299302(6)         &       55681.6123      &       21.6    &       0.15    &       65245   &       65004   &       102     &       11.979  \\
11953224        &       2MASS J19012670+5023067 &       19 01 26.705    &       +50 23 06.60        &               &       B9 V SiCr       &       N       &       4.1596468(7)         &       55685.5986      &       35.3    &       0.29    &       51492   &       51349   &       97      &       13.255  \\
\hline
\multicolumn{15}{l}{$^{1}$ \citet{RM09} $\;$ $^{2}$An alternative classification in \citet{gray16} identifies this star as A9 V Eu. This, however, corresponds to the rather low Sloan $g$ band S/N ratio of 21. We do not find evidence for the presence of Eu in the corresponding spectrum.} \\
\multicolumn{15}{l}{$^{3}$ \citet{grenier99} $^{4}$\citet{macrae52} $^{5}$ \citet{murphy69}} \\
\end{tabular}
\end{adjustbox}                                                                                                                                                                 
\end{center}                                                                                                                                                                    
}                                                                                                                                                                       
\end{sidewaystable*} 

\begin{table}
\caption{Astrophysical parameters of our target stars, which have been extracted from \citet{mathur17} and employed in the construction of the Kiel diagrams shown in Fig. \ref{fig_HRD}, sorted by KIC number. Columns are explained in the text (Sect. \ref{result}). We note that no parameters from this catalogue are available for KIC\,4136285.}
\label{table_data2}
\scriptsize{
\begin{center}
\begin{adjustbox}{max width=\textwidth}
\begin{tabular}{lllllll}
\hline
\hline
Name    &       \teff\  &       \teff\ err$+$   &       \teff\ err$-$   &       \logg\  &       \logg\ err$+$  &       \logg\ err$-$   \\
\hline
1576144 &       9181    &       251     &       430     &       4.003   &       0.234   &       0.156   \\
1865567 &       9888    &       275     &       412     &       4.209   &       0.144   &       0.246   \\
2441702 &       7828    &       218     &       354     &       3.791   &       0.337   &       0.112   \\
2853320 &       11357   &       587     &       1762    &       3.667   &       0.448   &       0.112   \\
2969628 &       8241    &       228     &       370     &       3.722   &       0.428   &       0.143   \\
3326428 &       8590    &       235     &       404     &       3.981   &       0.216   &       0.144   \\
3561656 &       9744    &       272     &       409     &       4.123   &       0.148   &       0.222   \\
3945892 &       8041    &       224     &       352     &       4.183   &       0.065   &       0.208   \\
%4136285        &       n/a       &     n/a     &       n/a     &       n/a       &      n/a       &     n/a     \\
4828345 &       11617   &       744     &       1116    &       3.666   &       0.552   &       0.097   \\
5213466 &       8527    &       235     &       369     &       3.765   &       0.412   &       0.165   \\
5473826 &       11063   &       348     &       502     &       3.999   &       0.266   &       0.143   \\
5727964 &       8124    &       226     &       340     &       4.065   &       0.165   &       0.135   \\
5739204 &       9099    &       251     &       466     &       4.065   &       0.181   &       0.148   \\
5774743 &       8203    &       228     &       342     &       3.739   &       0.413   &       0.110   \\
5814635 &       7904    &       216     &       351     &       4.130   &       0.098   &       0.182   \\
6065699 &       11063   &       353     &       530     &       3.999   &       0.266   &       0.143   \\
6206125 &       9591    &       378     &       648     &       3.602   &       0.476   &       0.084   \\
6278403 &       10932   &       228     &       495     &       4.056   &       0.236   &       0.193   \\
6305572 &       9345    &       294     &       425     &       4.140   &       0.099   &       0.231   \\
6310414 &       9643    &       272     &       409     &       4.142   &       0.148   &       0.222   \\
6426158 &       10155   &       286     &       429     &       4.183   &       0.151   &       0.280   \\
6715809 &       8287    &       231     &       346     &       4.001   &       0.204   &       0.119   \\
6864569 &       11076   &       309     &       530     &       3.989   &       0.253   &       0.156   \\
6950556 &       11119   &       398     &       531     &       3.642   &       0.425   &       0.075   \\
7628336 &       7478    &       233     &       285     &       3.859   &       0.400   &       0.100   \\
7778838 &       11287   &       612     &       1716    &       3.642   &       0.476   &       0.084   \\
8012437 &       8319    &       230     &       374     &       3.735   &       0.428   &       0.143   \\
8161798 &       10974   &       228     &       495     &       4.041   &       0.252   &       0.168   \\
8324268 &       9265    &       222     &       381     &       4.200   &       0.062   &       0.248   \\
8362546 &       11430   &       364     &       445     &       4.354   &       0.072   &       0.144   \\
8415109 &       9709    &       306     &       409     &       4.024   &       0.225   &       0.184   \\
8569986 &       9176    &       255     &       438     &       3.929   &       0.270   &       0.180   \\
8619436 &       10894   &       210     &       506     &       3.667   &       0.368   &       0.092   \\
8773445 &       8359    &       231     &       363     &       3.742   &       0.428   &       0.143   \\
8881883 &       10949   &       221     &       516     &       3.650   &       0.416   &       0.073   \\
9541567 &       9274    &       255     &       475     &       4.092   &       0.158   &       0.193   \\
10082844        &       9643    &       272     &       428     &       4.142   &       0.167   &       0.204   \\
10281890        &       10906   &       220     &       515     &       4.063   &       0.200   &       0.200   \\
10324412        &       10153   &       318     &       424     &       3.543   &       0.532   &       0.028   \\
10550657        &       9274    &       254     &       471     &       4.092   &       0.170   &       0.170   \\
10685175        &       8709    &       462     &       925     &       4.232   &       0.032   &       0.288   \\
10905824        &       9354    &       263     &       489     &       4.061   &       0.171   &       0.209   \\
10959320        &       9824    &       272     &       408     &       3.750   &       0.322   &       0.138   \\
10971633        &       9189    &       286     &       430     &       3.952   &       0.264   &       0.176   \\
10982373        &       9453    &       399     &       699     &       3.832   &       0.185   &       0.203   \\
11098975        &       9842    &       272     &       443     &       4.213   &       0.136   &       0.253   \\
11154043        &       8912    &       249     &       427     &       4.050   &       0.181   &       0.165   \\
11465134        &       9859    &       274     &       445     &       4.166   &       0.160   &       0.240   \\
11560273        &       9225    &       251     &       466     &       3.783   &       0.399   &       0.171   \\
11671226        &       8300    &       201     &       374     &       3.744   &       0.432   &       0.135   \\
11807603        &       8814    &       246     &       422     &       4.099   &       0.135   &       0.165   \\
11953224        &       9828    &       275     &       412     &       4.106   &       0.126   &       0.234   \\
\hline
\end{tabular}
\end{adjustbox}                                                                                                                                                                 
\end{center}                                                                                                                                                                    
}                                                                                                                                                                       
\end{table}

%\clearpage

%-----------------------------------------------------------------------

\section{Discussion} \label{discussion}

In the following discussion concerning the light curve properties of ACV variables, we only considered objects that have been spectroscopically confirmed as mCP stars ($N$\,=\,41). We caution that the present sample, which constitutes the largest collection of mCP stars in the \kepler\ field to date, is still rather small and biased by the employed selection criteria. It is thus not very well suited to statistical analyses. It is, however, sufficiently large and adequate for an investigation of the goals of the present study.

\subsection{mCP star light curve properties} \label{photvar}

\begin{figure}
        \includegraphics[width=\columnwidth]{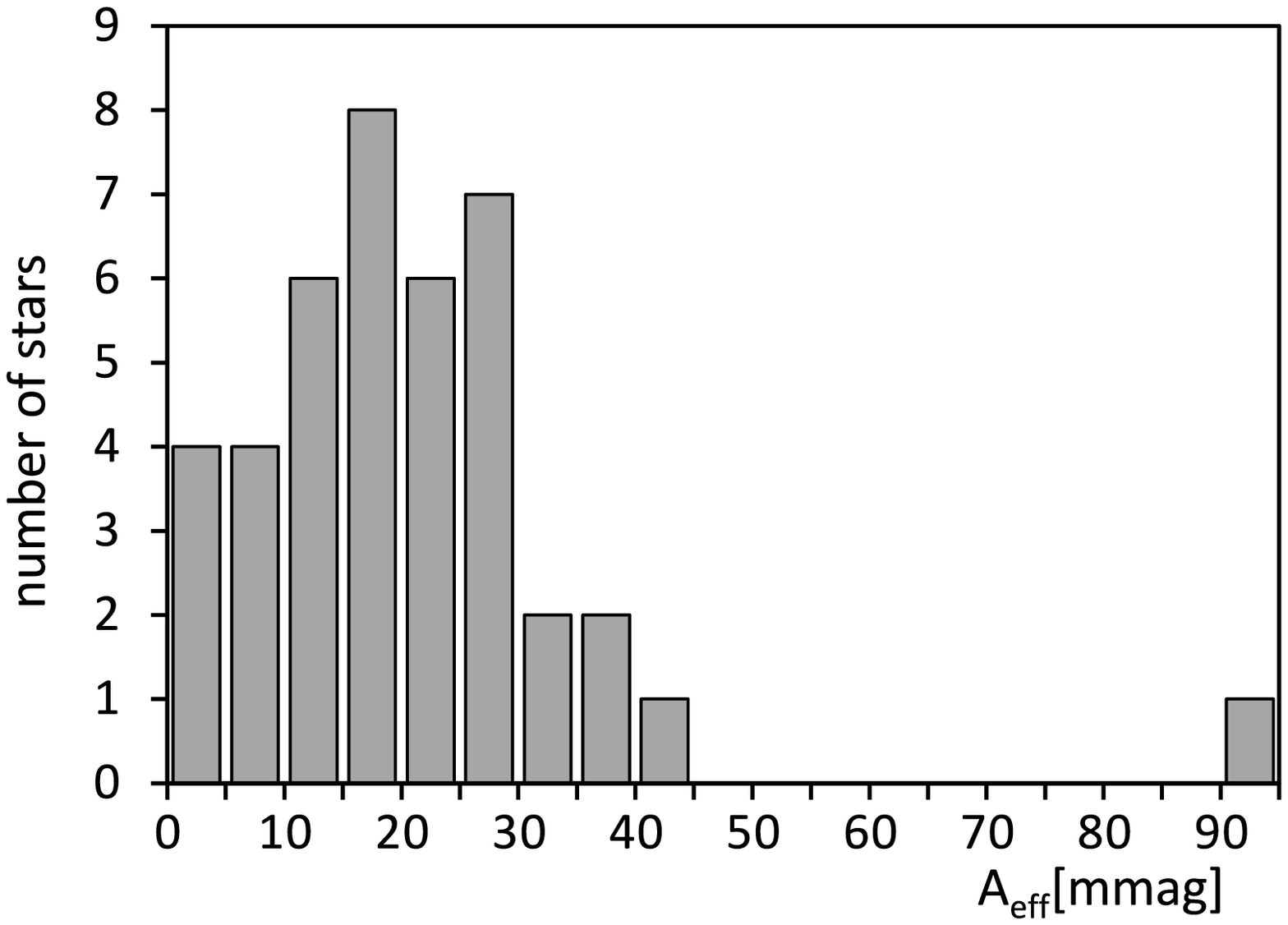}
    \caption{Distribution of amplitudes among the spectroscopically confirmed ACV variables in our sample ($N$\,=\,41).}
    \label{fig_amplitudes}
\end{figure}

The distribution of effective amplitudes of our sample stars is shown in Fig. \ref{fig_amplitudes}. Our results are in good agreement with the literature. ACV variables that have been detected in ground-based photometric studies mostly exhibit amplitudes of the order of several hundredths of a magnitude \citep{mathys85}, peaking at around 30\,mmag (cf. lower panel of Figure 1 in \citealt{bernhard15a}). In accordance with our expectations, \kepler\ data yield a significant number of ACV variables with very small amplitudes. This is expected as the range of the light variations depends on the contrast and size of the involved chemical spots that may, at least in theory, assume all possible configurations. The star KIC\,8161798 (A0 V SiCr) deserves special note, as it shows by far the largest effective amplitude (90.5\,mmag) of our sample stars.

\begin{figure}
        \includegraphics[width=\columnwidth]{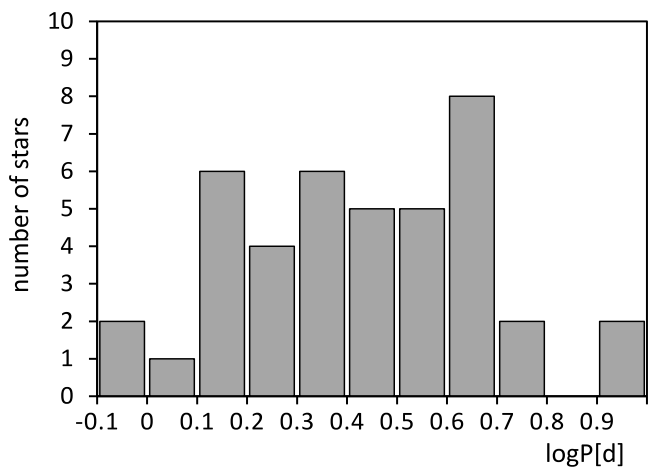}
    \caption{Distribution of periods among the spectroscopically confirmed ACV variables in our sample ($N$\,=\,41).}
    \label{fig_periods}
\end{figure}

The distribution of periods is illustrated in Fig. \ref{fig_periods}. It is typical of ACV variables, which show a peak distribution around a period of log\,$P$\,$\sim$\,0.4, where $P$ is expressed in days (cf. Figure 1 in \citealt{bernhard15a} and Figure 7 in \citealt{RM09}). At around a rotational period of log\,$P$\,$\sim$\,0.8, the incidence of ACV variables generally decreases. However, we note that, as has been pointed out in Section \ref{sample_selection}, this representation is biased by our sample selection criteria, which are bound to miss long period ACV variables. Furthermore, the small sample size precludes drawing any definite conclusions.

The availability of precise light curves with a maximum of information content is important to derive accurate information on spot sizes and distribution and atmospheric structure in mCP stars. \kepler\ data are used for the first precise characterizations of the light curves of a larger sample of these stars and open up a new window on their rotational light variability properties. While the majority of mCP stars in our sample show rather simple light changes that can indeed be well represented by a sine wave and its first harmonic, about 25\,\% of these objects show a wealth of details in their light curves that hint at complex surface structures. This is in particular the case for the stars KIC 4136285, KIC 5774743, KIC 5814635, and KIC 6426158 (cf. Fig. \ref{allphased}).

\begin{figure*}[h!]
 \centering
%        \mbox{
%                        \subfigure{\includegraphics[width=0.47\textwidth]{scatterplot1_1.eps}}
%                        \subfigure{\includegraphics[width=0.47\textwidth]{scatterplot1_2.eps}}
%        } 
%        \mbox{
%                        \subfigure{\includegraphics[width=0.47\textwidth]{scatterplot1_3.eps}}
%                        \subfigure{\includegraphics[width=0.47\textwidth]{scatterplot1_4.eps}}
%        }
%        \mbox{
%                        \subfigure{\includegraphics[width=0.47\textwidth]{scatterplot3_1.eps}}
%                        \subfigure{\includegraphics[width=0.47\textwidth]{scatterplot3_2.eps}}
%        } 
	      \mbox{
                        \includegraphics[width=0.47\textwidth]{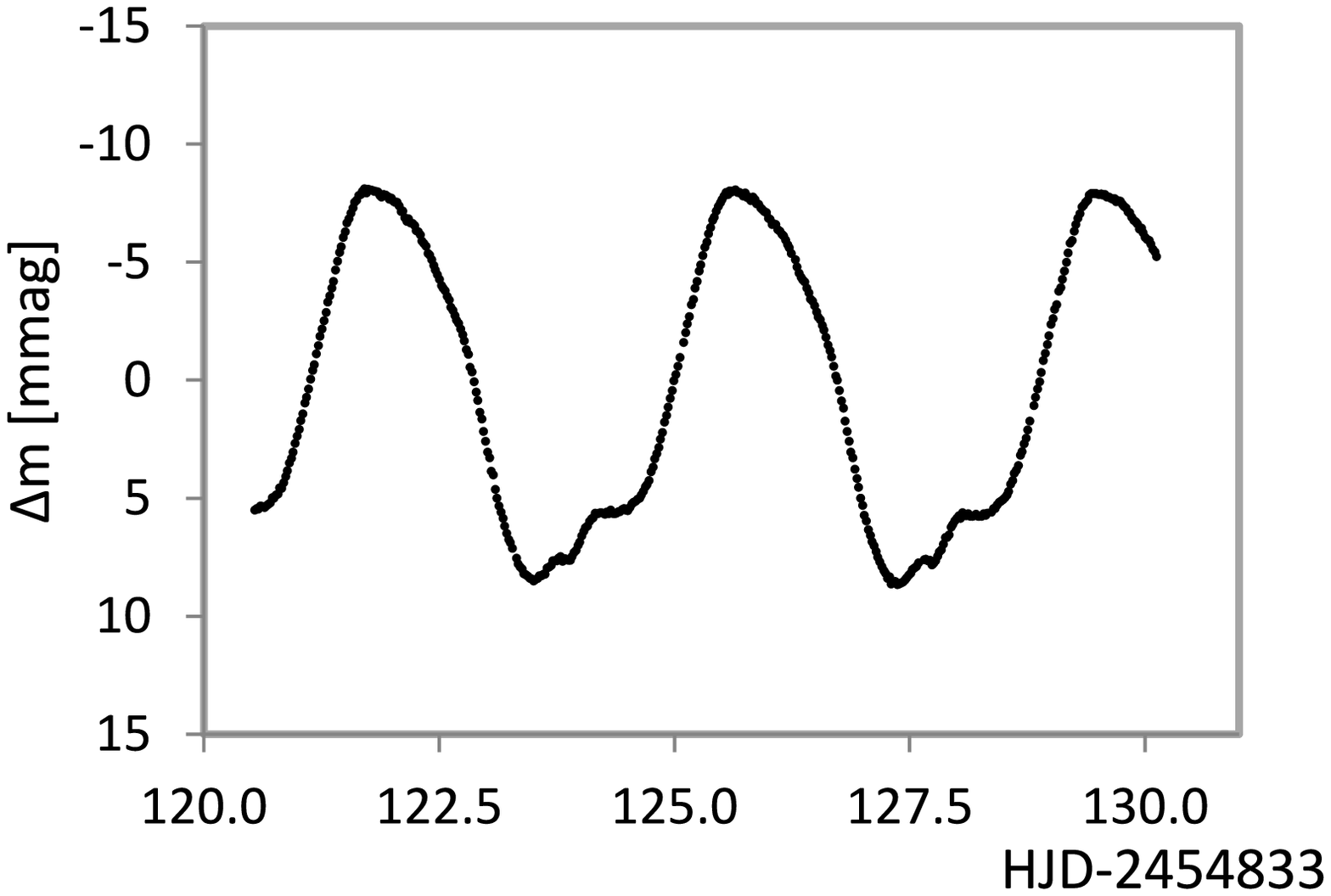}
                        \includegraphics[width=0.47\textwidth]{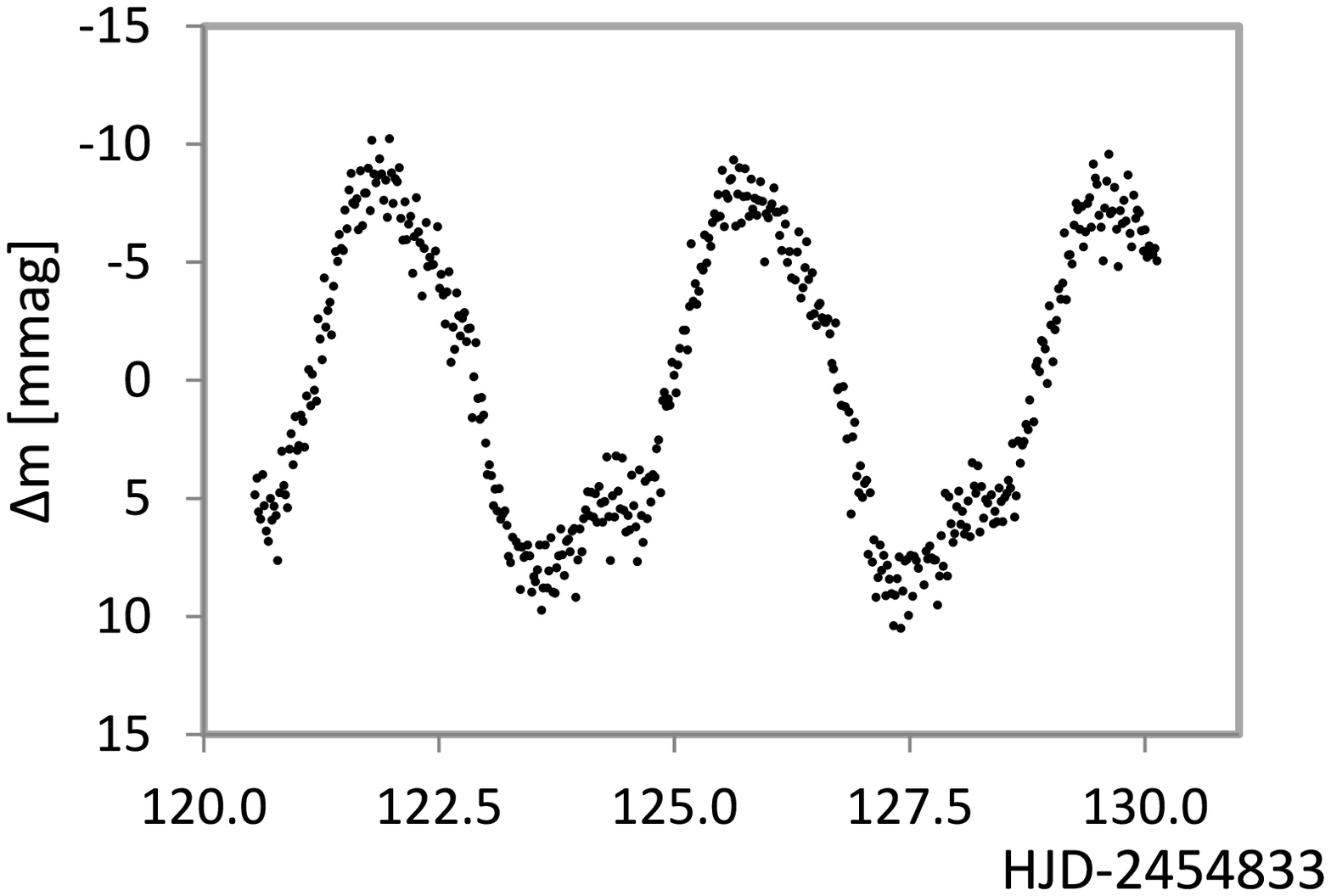}
        } 
        \mbox{
                        \includegraphics[width=0.47\textwidth]{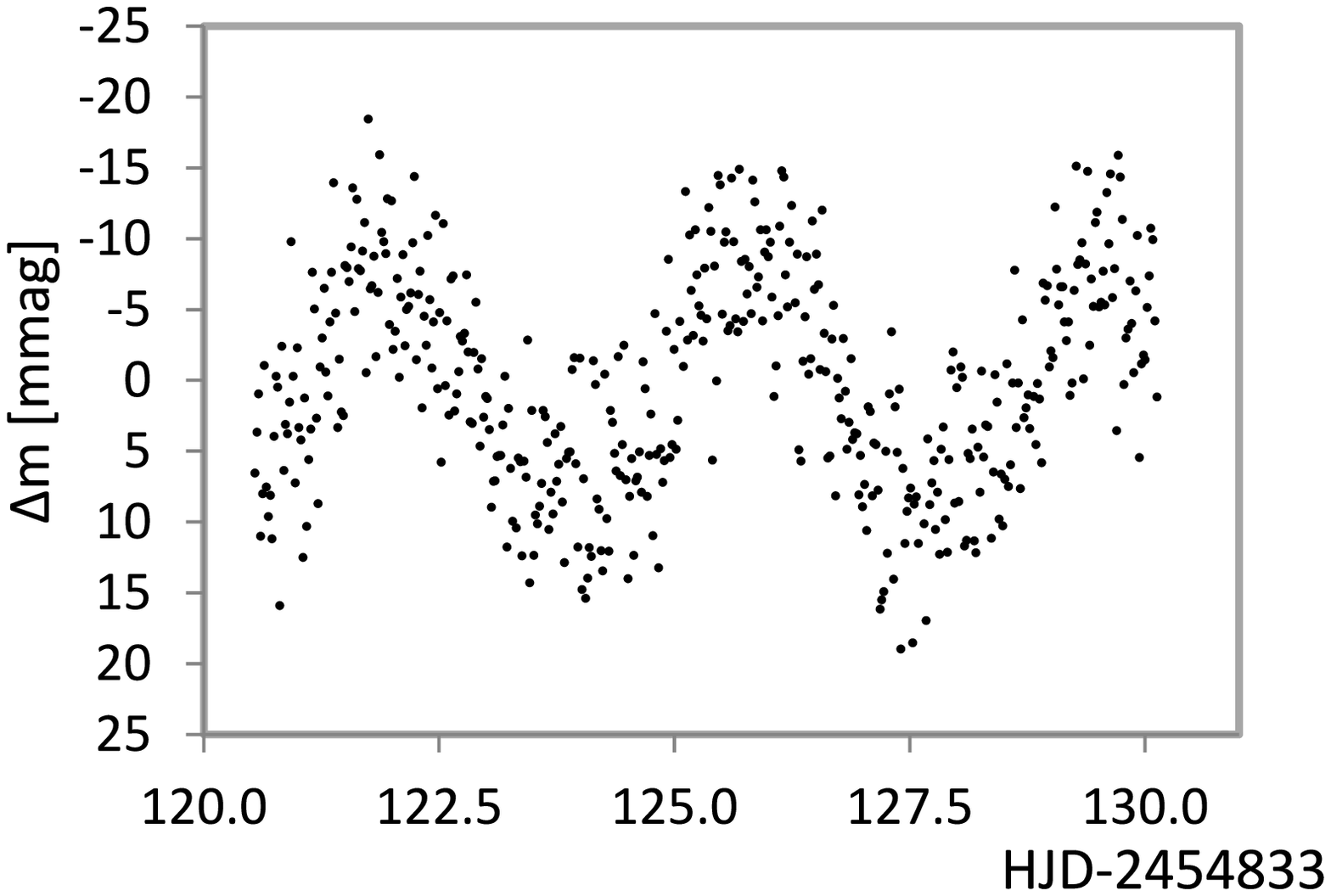}
                        \includegraphics[width=0.47\textwidth]{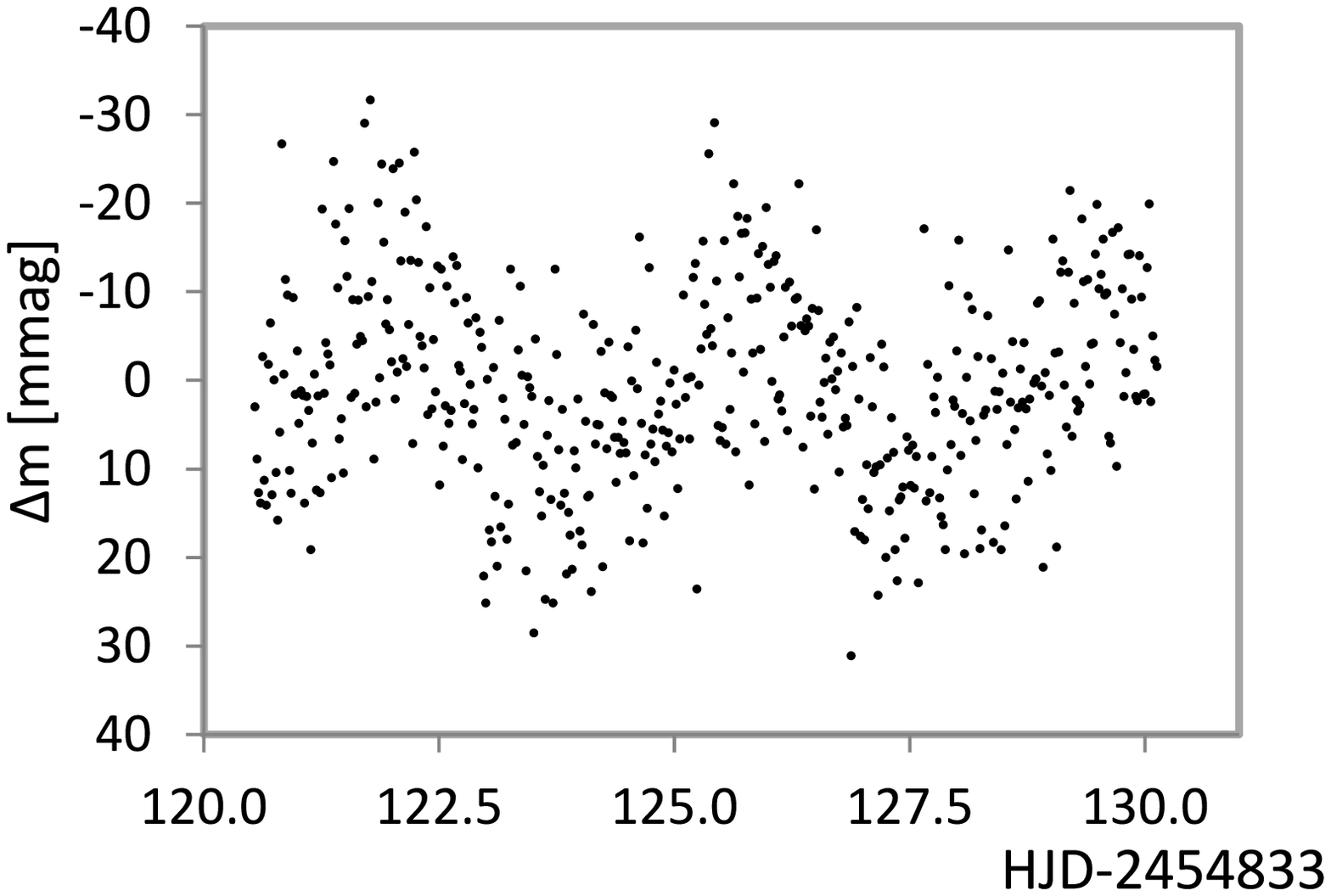}
        }
        \mbox{
                        \includegraphics[width=0.47\textwidth]{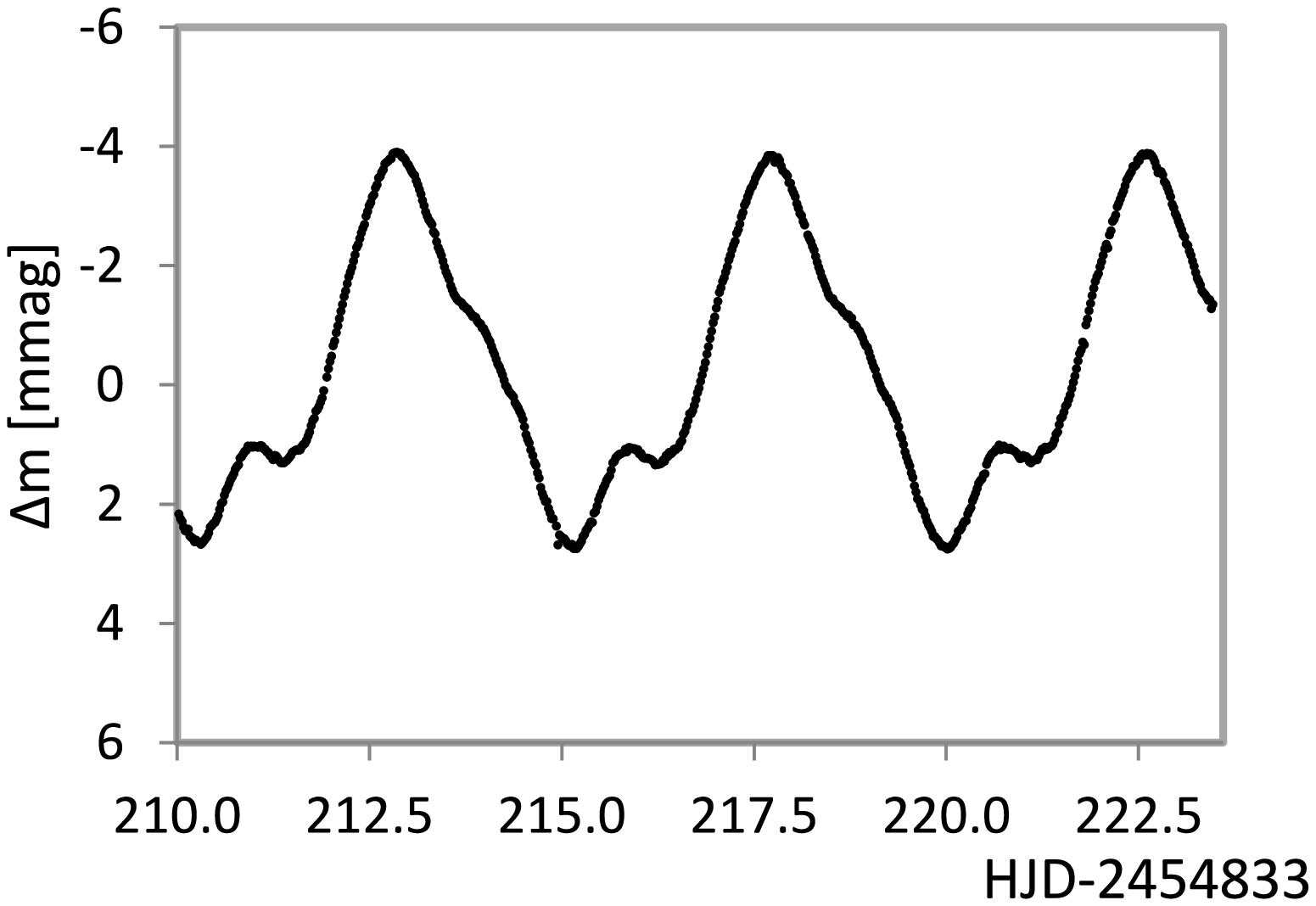}
                        \includegraphics[width=0.47\textwidth]{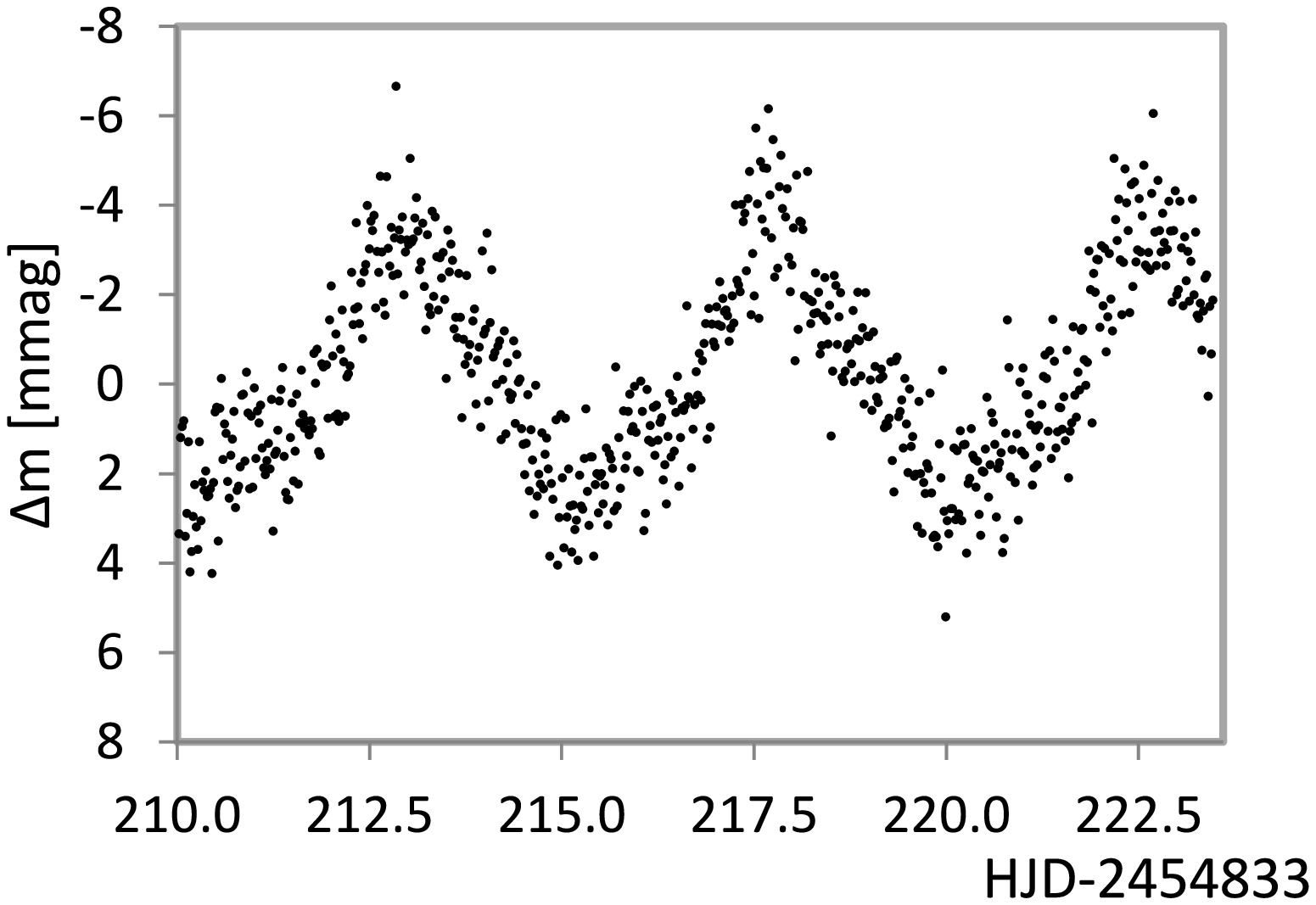}
        } 
 \caption[]{Light curve of KIC\,5814635 (A1 V SiSrCrEu metals strong) based on original \kepler\ data (upper left panel) and \kepler\ data to which a Gaussian scatter of 1\,mmag (upper right panel), 5\,mmag (middle left panel), and 10\,mmag (middle right panel) has been added. The subtle details revealed by \kepler\ can only be guessed after the introduction of a scatter of 1\,mmag, which become totally unrecognizable as the scatter increases to 5 and 10\,mmag, respectively. The same holds true for KIC\,6426158 (A1 V SiSrCrEu). The details in the light curve (lower left panel) already get blurred after the introduction of a scatter of 1\,mmag (lower right panel). This illustrates that the subtle details revealed by \kepler\ would have been lost in standard ground-based observations.}
 \label{scatterplot1}
\end{figure*}

In order to check whether these details would be detectable in less accurate observations typical of ground-based data, we introduced Gaussian scatter into the light curves of several objects (Fig. \ref{scatterplot1}). As can be clearly seen, the subtle details in the light curve of KIC 5814635 can only be guessed after the introduction of a scatter of 1\,mmag, becoming totally unrecognizable as the scatter increases to 5 and 10\,mmag, respectively. The same holds true for the low-amplitude ACV variable KIC 6426158, in which the subtle light curve details are already lost after the introduction of a scatter of 1\,mmag.

These examples illustrate that, at least in some mCP stars, high-precision photometry reveals previously undetected light curve details, which indicate that the surface structures of these stars are more complex than hitherto observed. In this context, it is interesting to point out that several investigations have reported complex magnetic structures in mCP stars, for example HD 32633 \citep{renson84}, $\beta$ CrB \citep{bagnulo00}, 53\,Cam \citep{kochukhov04}, and 49\,Cam \citep{silvester17}. With increasing quality of observations, it has become clear that most mCP stars have fields that are considerably more complex than pure oblique dipoles \citep{bagnulo02}. As magnetic and surface abundance structures are related \citep{kochukhov_thesis}, these findings imply the existence of complex surface abundance structures, and hence complex photometric variability, in at least some mCP stars; this is in agreement with our results from \kepler\ data. It would be highly interesting to establish whether the complex photometric variations in our sample stars are reflected in the magnetic field curves, and we strongly encourage corresponding DI studies. We cannot answer the question concerning which other parameters (if any) determine the occurrence of complex versus simple surface structures in mCP stars with the current sample, although it is, of course, reasonable to assume a continuous progression from simple to complex surface structures among these objects. It may, however, be tackled in the future when a larger sample of mCP stars with ultra-precise light curves has become available.

Another noteworthy result of our analysis is that, apparently, the spectroscopically confirmed mCP stars indeed show stable light curves during the full period of the \kepler\ coverage. This is in accordance with the expectations, as ACV variables are believed to exhibit stable spot configurations (and hence light curves) during long periods of time (cf. Sect. \ref{introduction}). On the other hand, all stars for which we failed to establish chemical peculiarities show some kind of long-term changes in light curve shape (in particular KIC 5213466, KIC 10082844, and KIC 10550657) or some kind of additional variability (cf. Sect. \ref{nonCPs}). This apparent dichotomy might prove valuable in future efforts aimed at the discovery of mCP star candidates by their light curve properties
in large photometric surveys. An automated approach seems readily conceivable and is under consideration. However, we caution that the differentiation between instrumental effects and intrinsic light curve features in \kepler\ data is not easy.

\subsection{Objects of special interest} \label{nonCPs}

For seven stars of our sample, we were not able to detect peculiarities in the available spectroscopic material. None of these objects show a 5200\,\AA\ flux depression (Table \ref{table_data1}) or are outstanding in the Kiel diagram (Fig. \ref{fig_HRD}). This raises the question of whether the spectral peculiarity is too weak to be detected in our low-resolution spectroscopic material or whether the variability is not due to chemical spots on the surface. The fact that these objects do not exhibit stable light curves during the \kepler\ coverage clearly favours the latter explanation.

Several studies have reported a surprisingly high rate of rotational variables among early-type (B/A) main-sequence stars \citep[e.g.][]{balona13b,balona15,balona17}, and there have been speculations about the presence of localized magnetic fields and "classical" starspots in these stars \citep{balona17}. This, however, is still a controversial issue \citep[e.g.][]{murphy14}. In a recent study, \citet{saio18} presented theoretical calculations and evidence for the occurrence of global Rossby waves (r modes) in upper main-sequence stars. Assuming swarms of r-mode frequencies and weak starspots, they have been able to reproduce the observed Fourier amplitude spectra of rapidly rotating B/A stars, thus producing independent evidence for the occurrence of classical starspots in these objects.

Our non-mCP stars exhibit light curves that are compatible with spot-induced variability. Furthermore, the occurrence of long-term changes or some kind of additional variability in these stars suggests classical starspots rather than chemical spots, which are longer lived. Further investigations of these interesting targets may contribute to the resolution of this issue and are strongly encouraged. Only detailed spectroscopic time-series analyses will be able to shed more light on the true nature of these objects, which are briefly commented on below. Spectral types derived in this study are provided in parentheses behind the KIC identifiers.

KIC\,5213466 (A1 V), KIC\,10082844 (A0 V), and KIC\,10550657 (A0 V): The light curves of these stars show large variations in amplitude and shape (cf. Figs. \ref{development} and \ref{allphased}), which is not expected in ACV variables. The reason could be differential rotation, binary motion, or starspots due to stellar activity. Interestingly, the spectral types and light curves of all three stars are very similar.

KIC\,5727964 (A6 V): This star shows alternating heights of consecutive maxima, which is clearly observed in \kepler\ data (cf. Fig. \ref{allphased}) but already lost after the introduction of a scatter of 1\,mmag. Therefore, in less accurate data, only half the true period would have been registered. The light curve is continually (but marginally) changing throughout the covered time span.

KIC\,6206125 (B9 V (Si)): The periodogram indicates the presence of additional short-period peaks. The most significant peak is at $f_1$\,=\,1.88555\,d$^{-1}$ ($P$\,=\,0.53035\,d) and has an amplitude of only $\sim$0.2\,mmag. This additional variability is present in all quarters, and we do not see any evidence that it is caused by instrumental effects. In fact, the star has been proposed as a $\delta$ Scuti variable by \citet{balona13b} on grounds of the presence of high-frequency peaks and is listed as such in SIMBAD, although the observed main period seems to be rather long for this type of variable star. We confirm the Si peculiarity indicated by \citet{gray16} even though the star is a marginal case. The possible presence of noAp (i.e. $\delta$ Scuti or $\gamma$ Doradus) pulsations in CP2 stars have been reported in the literature \citep[e.g.][]{balona11b}, and the simultaneous existence of $\delta$ Scuti pulsations and rotational modulation due to an inhomogeneous surface structure has been established in HD 41641, which also shows the spectral characteristics of a CP2 star \citep{escorza16}. These objects are rare, therefore KIC\,6206125 is potentially of interest and merits further attention. However, the star has been identified as a binary star \citep{murphy18}, and it needs to be excluded that the short period signal originates in the companion star or that \kepler\ photometry is contaminated by the signal of a nearby pulsating variable in the same field of view.

KIC\,8415109 (A0 V): The amplitude of the light variations is extremely small (2.3\,mmag). The light curve is stable and its complex pattern is suggestive of spot-induced variability. This is an interesting target for follow-up observations because this source could be a very young CP star in the process of developing chemical peculiarities.

KIC 8569986 (A2 V) and KIC 11671226 (A5 V): No outstanding characteristics compared to the confirmed ACV variables are visible.

Two other objects deserve special mention and are briefly discussed below.

\begin{figure}
\begin{center}
        \includegraphics[width=0.9\columnwidth]{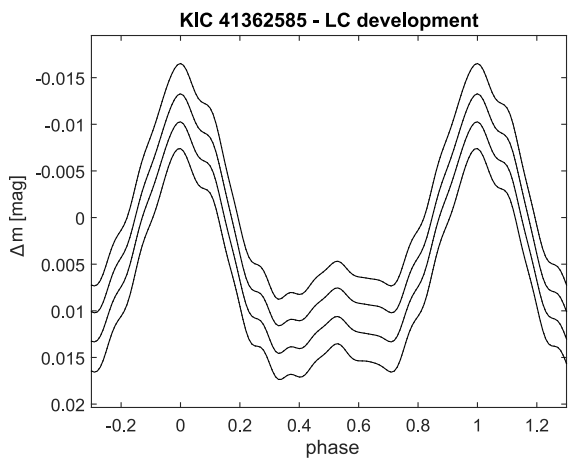}
    \caption{Light curve development of KIC\,4136285 (spectral type B5 He wk.; \citealt{RM09}). The plot has been created with the DEVELOPMENT code and based on original, non-detrended Kepler PDC flux. The time distance between subsequent mean light curves is about 45 days. The star exhibits a highly complex but stable light curve throughout the covered time span.}
    \label{development3}
\end{center}
\end{figure}

KIC\,4136285 (B5 He wk., \citealt{RM09}): This star, which is also known as V545\,Lyr, is the only CP4 star in our sample. It exhibits a highly complex light curve, which is stable throughout the covered time span (Fig. \ref{development3}). We interpret the light variations as rotational modulation caused by abundance inhomogeneities. The observed variability pattern is therefore indicative of a highly complex surface structure, and the object is a prime target for detailed spectroscopic analyses or DI studies. Unfortunately, only two quarters of data are available for this interesting star.

KIC\,6950556 (A0 V Si): This star has been classified as A8mB4 V Lam Boo by \citet{gray16}. The analysed spectrum boasts a Sloan $g$ band S/N value of 145, which should allow for a precise classification. However, we classified the star as A0 V Si in the present investigation. A strong 5200\,\AA\ flux depression is present in the spectrum that is characteristic of mCP stars and provides confidence in our classification. This star is certainly not a $\lambda$ Bootis star.

%-----------------------------------------------------------------------

\section{Conclusions} \label{conclusion}

We have carried out a search for new photometrically variable mCP stars (ACV variables) in the \kepler\ field with the ultimate aim of investigating their photometric variability properties. For the creation of an initial sample, photometric selection criteria were applied to the \kepler\ light curves of early-type targets, which lead to the identification of 51 photometric candidates. These candidates, together with two well-confirmed mCP stars (KIC 4136285 and KIC 8324268), formed the final sample for the present study. Employing newly acquired and archival spectra, we were able to investigate 46 photometric candidates, 39 of which (85\%) turned out to be bona fide mCP stars.

The main findings of the present investigation are summarized in the following:

\begin{itemize}
        \item Our final sample consists of 41 spectroscopically confirmed mCP stars, 39 of which are new discoveries; 5 candidate mCP stars for which no (recent) spectra are available; and 7 stars in which no chemical peculiarities could be established. We hereby significantly enlarge the sample of known mCP stars in the \kepler\ field and provide a sound basis for further research.
        \item Observed periods range from 0.84 d to 9.6 d, and effective amplitudes range from 0.6 mmag to 90.5 mmag.
        \item The employed photometric selection criteria constitute a viable and efficient method of identifying mCP star candidates among early-type stars in massive photometric time series databases. Light curve stability, in particular, has been identified as a primary criterion in this respect.
        \item Our targets populate the whole age range from ZAMS to terminal-age main sequence (TAMS) and are distributed in the mass interval from 1.5\,$M_{\odot}$ to 4\,$M_{\odot}$. No evolved objects or stars below the ZAMS are present in our sample. The derived spectral types are in good agreement with the \teff\ values of \citet{mathur17}.
        \item About 25\,\% of the \kepler\ mCP stars show a hitherto unobserved wealth of detail in their light curves that is indicative of complex surface structures. These data are valuable for detailed analyses of the abundance inhomogeneities and atmospheric structure in mCP stars and provide important input parameters for spot-modelling attempts and simulation of mCP star light curves.
        \item For 7 stars, no evidence for chemical peculiarities has been found. These stars, however, show variability that is compatible with rotational modulation and are interesting targets for follow-up investigations.
        \item We provide independent confirmation of the accuracy of the spectral classifications derived by the MKCLASS code, in particular concerning the reliability of the peculiarity classifications. %The MKCLASS code spectral types are shown to form a reliable basis for further investigations of astrophysically interesting objects.
\end{itemize}

Future investigations will be concerned with the identification of more mCP stars in the \kepler\ field via photometric selection criteria and the MKCLASS code and a detailed analysis of the light variability of the more interesting objects of the present sample ({Mikul{\'a}{\v s}ek} et al., in preparation).

\begin{figure*}
\begin{center}
\includegraphics[width=1.0\textwidth]{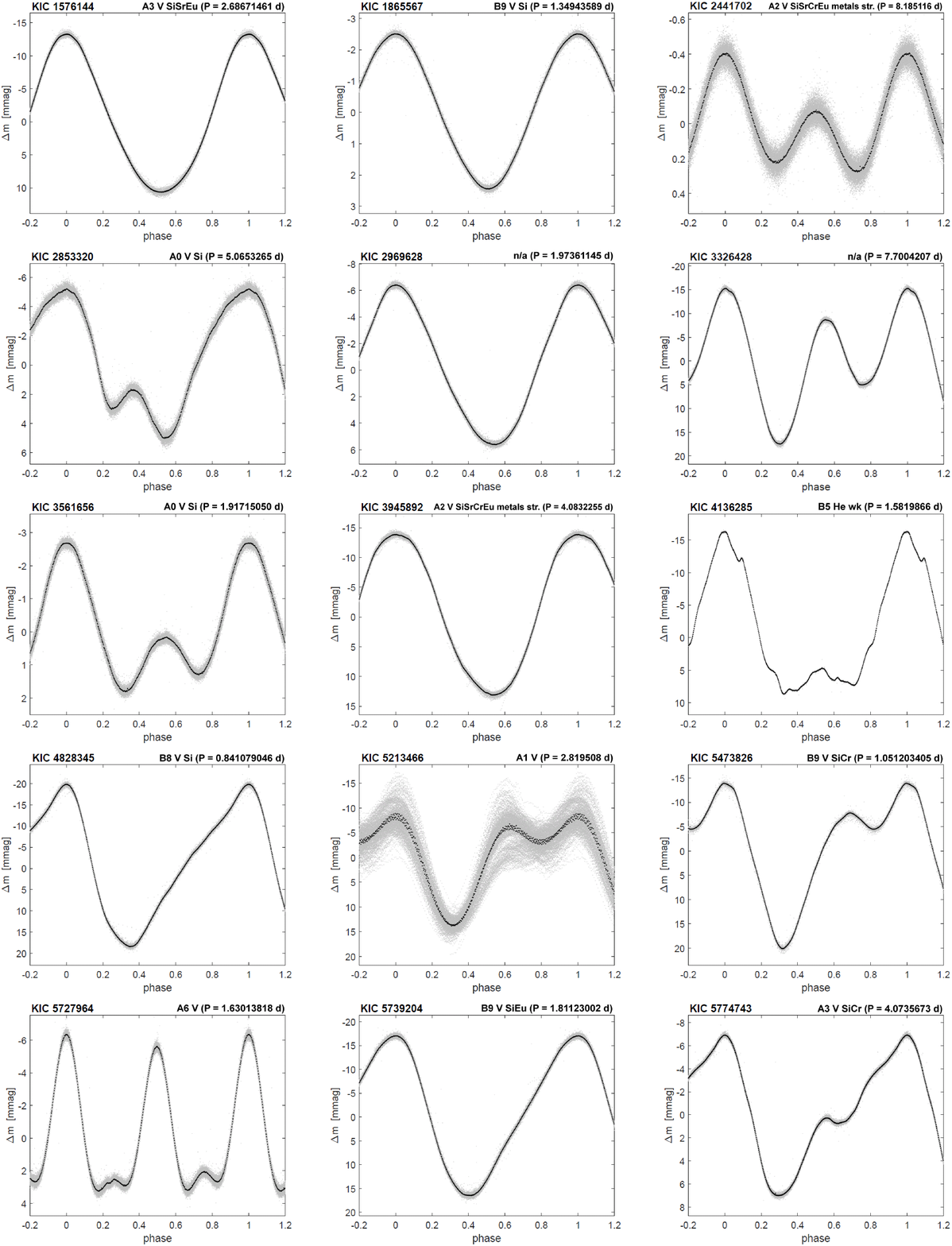}
\caption{Light curves of all objects, based on detrended and rescaled \kepler\ data (cf. Sect. \ref{ephemerides}) and phased with the linear ephemerides presented in Table \ref{table_data1}. The contours of the mean light curve are delineated by black dots. Phase zero corresponds to photometric maximum. The indicated periods and spectral types have been gleaned from Table \ref{table_data1}. The given spectral types have been taken from \citet{RM09} (KIC\,4136285), \citet{murphy69} (KIC\,11560273), and this study (all other stars).} 
\end{center}
\end{figure*}
\setcounter{figure}{11}
\begin{figure*}
\begin{center}
\includegraphics[width=1.0\textwidth]{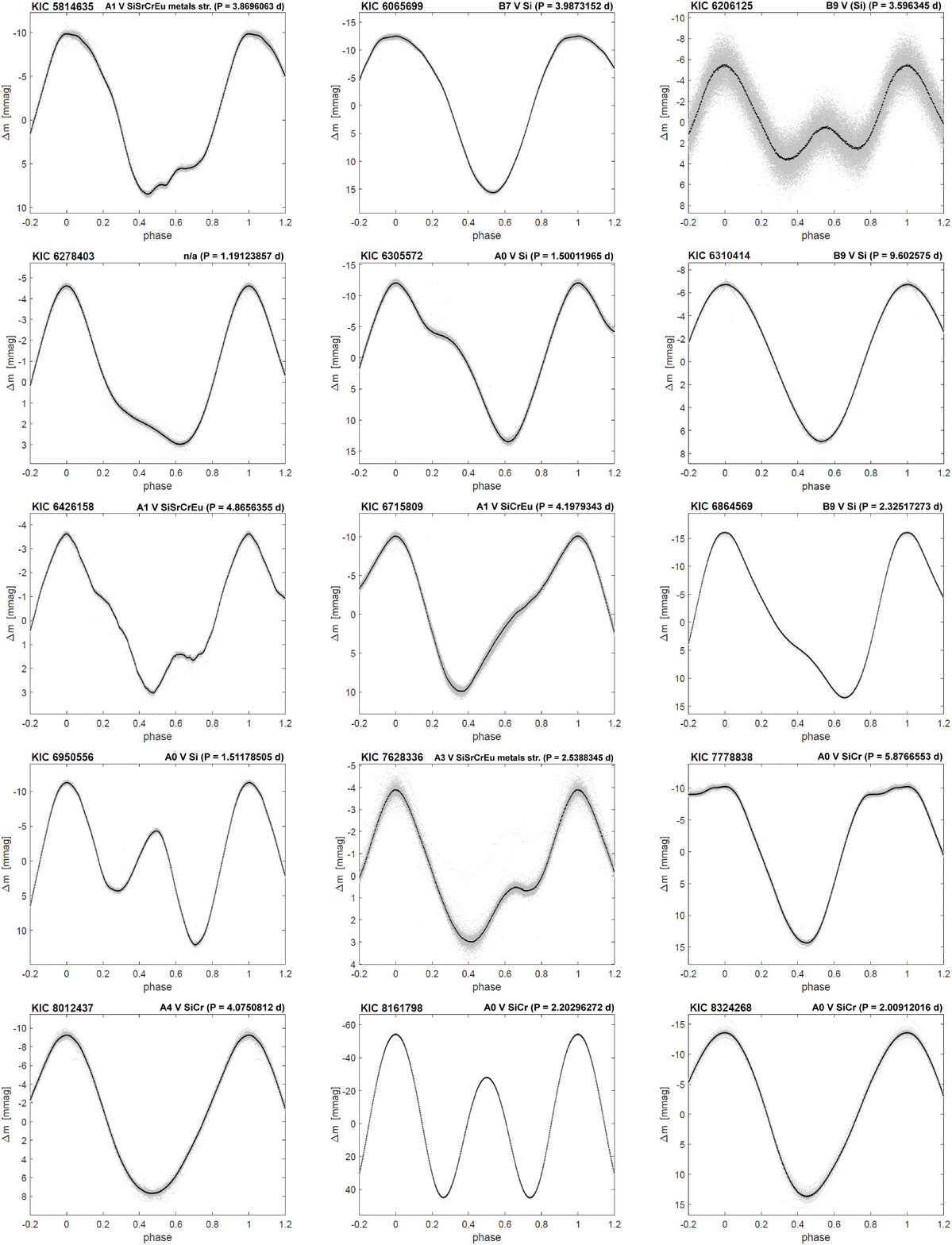}
\caption{continued.} 
\end{center}
\end{figure*}
\setcounter{figure}{11}
\begin{figure*}
\begin{center}
\includegraphics[width=1.0\textwidth]{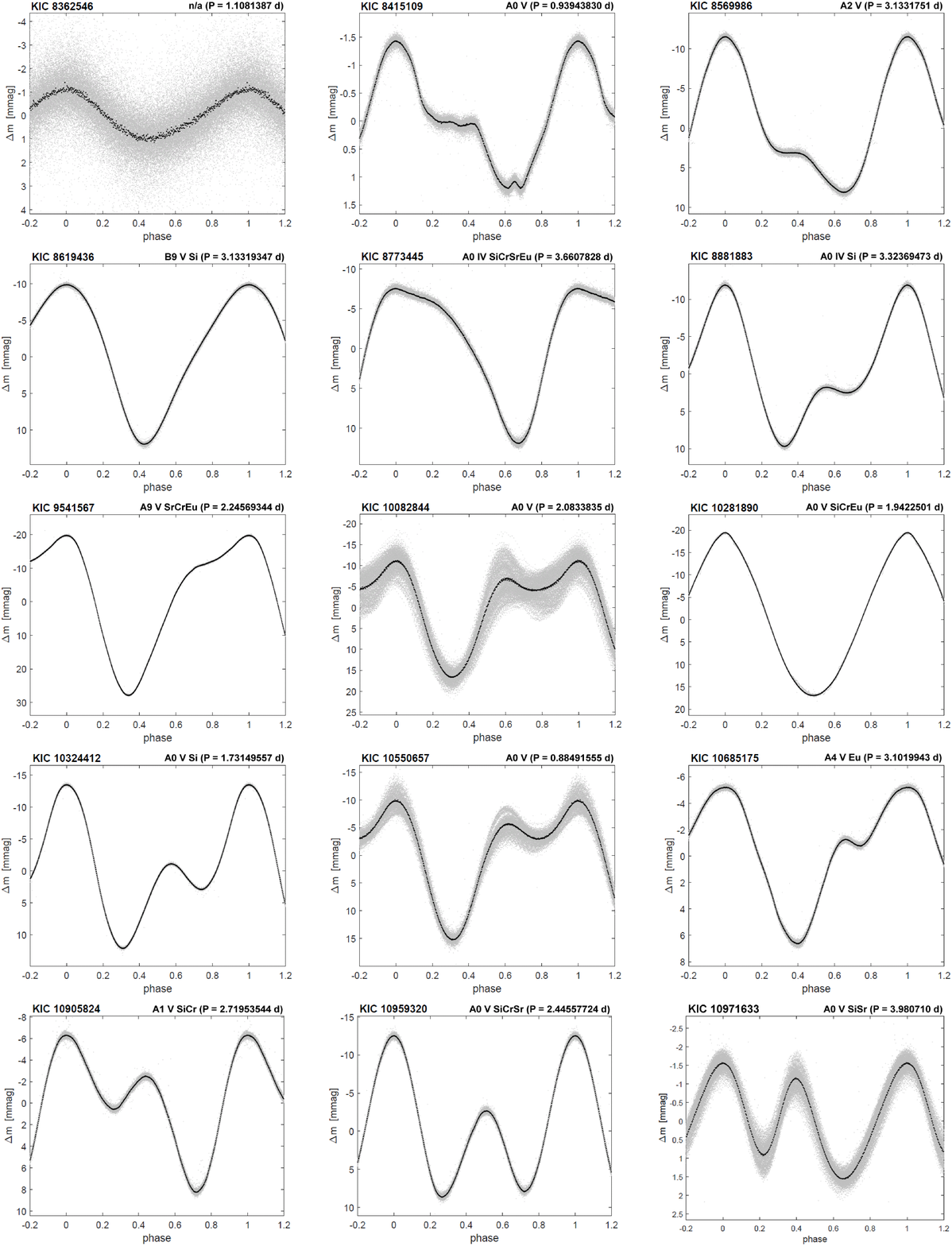}
\caption{continued.} 
\end{center}
\setcounter{figure}{11}
\end{figure*}\begin{figure*}
\begin{center}
\includegraphics[width=1.0\textwidth]{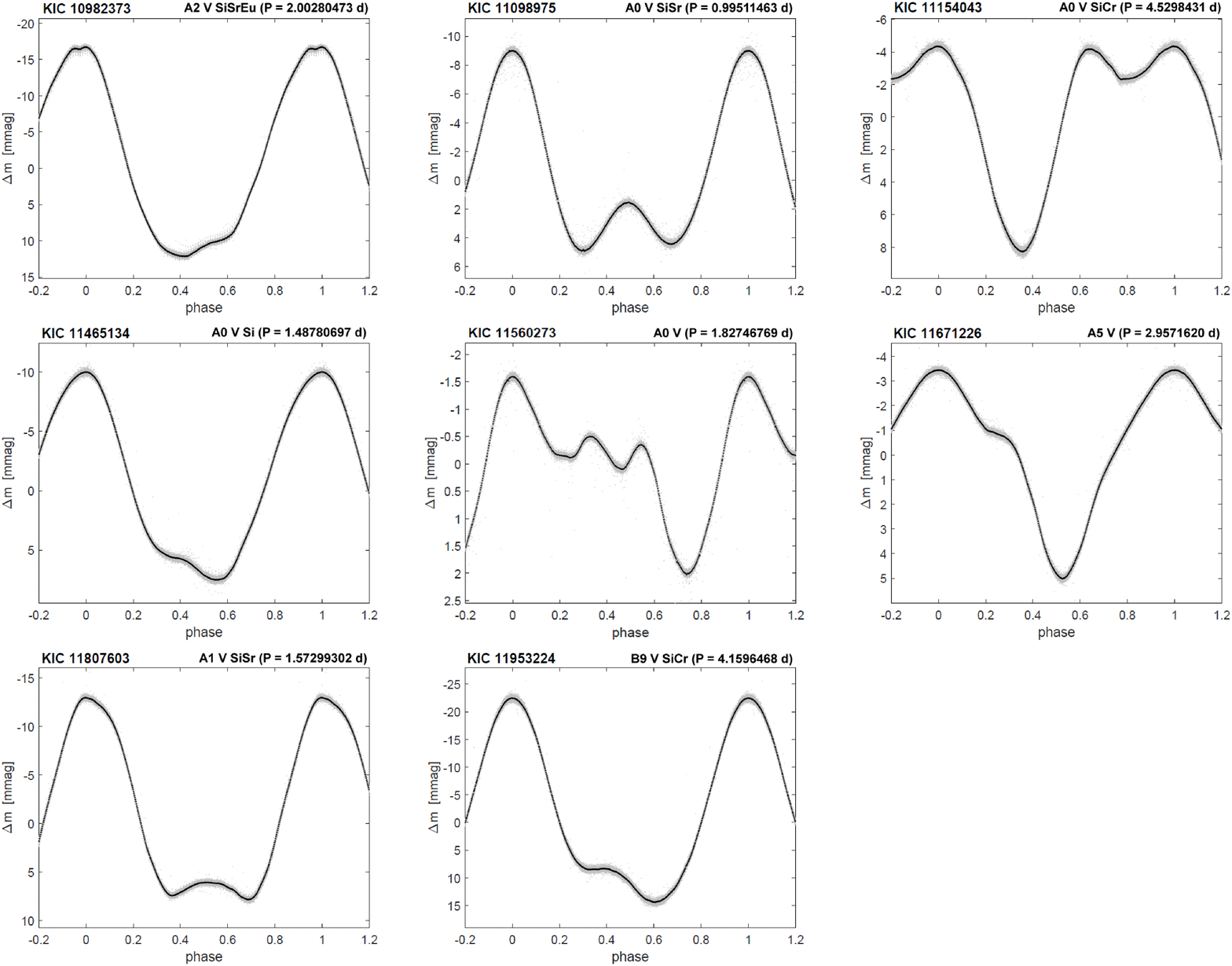}
\caption{continued.}
\label{allphased}
\end{center}
\end{figure*}

\begin{acknowledgements}
We thank the anonymous referee for the thoughtful report that helped to improve the paper. We also thank Prof. Don Kurtz for providing a list of candidate ACV variables that were included into the sample selection process. ZM acknowledges support from the GACR 16-011116S project. IY thanks the Russian Science Foundation for financial support (RSF grant 14-50-00043). This work has been funded by the projects VEGA 2/0031/18 and APVV-15-0458. This paper includes data collected by the \kepler\ mission. Funding for the Kepler mission is provided by the NASA Science Mission directorate. The \kepler\ data were obtained from the Mikulski Archive for Space Telescopes (MAST). STScI is operated by the Association of Universities for Research in Astronomy, Inc., under NASA contract NAS5-26555. Support for MAST for non-HST data is provided by the NASA Office of Space Science via grant NNX09AF08G and by other grants and contracts. Guoshoujing Telescope (the Large Sky Area Multi-Object Fiber Spectroscopic Telescope LAMOST) is a National Major Scientific Project built by the Chinese Academy of Sciences. Funding for the project has been provided by the National Development and Reform Commission. LAMOST is operated and managed by the National Astronomical Observatories, Chinese Academy of Sciences. This research has made use of the SIMBAD database and the VizieR catalogue access tool, operated at CDS, Strasbourg, France.
\end{acknowledgements}

\clearpage

%\begin{thebibliography}{}

\bibliographystyle{aa}
\bibliography{kepler1}

%\end{thebibliography}
%}
\end{document}